\documentclass[a4paper,fleqn,usenatbib]{mnras}

\usepackage{newtxtext,newtxmath}

\usepackage[T1]{fontenc}
\usepackage{ae,aecompl}
\usepackage[flushleft]{threeparttable}

\usepackage{graphicx}	
\usepackage{amsmath}	
\usepackage{amssymb}	

\usepackage{subfigure}

\title[The photocentre-AGN displacement in M87]{The photocentre-AGN displacement: Is M87 actually harbouring a displaced supermassive black hole?}

\author[E. L\'opez-Navas and M. A. Prieto ]{
E. L\'opez-Navas,$^{1}$\thanks{E-mail: elena.l.n.94@gmail.com}
and M. A. Prieto$^{1,2}$
\\
$^{1}$Departamento de Astrof\'isica, Universidad de La Laguna,E-38206 La Laguna, Tenerife, Spain\\
$^{2}$Instituto de Astrof\'isica de Canarias (IAC), E-38200 La Laguna, Tenerife, Spain}

\date{Accepted XXX. Received YYY; in original form ZZZ}

\pubyear{2018}

\begin{document}
\label{firstpage}
\pagerange{\pageref{firstpage}--\pageref{lastpage}}
\maketitle

\begin{abstract}
 M87 has been identified as a displaced supermassive black hole (SMBH) candidate. We investigated this possibility by a temporal analysis of twelve Adaptive-Optics assisted-VLT and HST images spanning twenty years. We found that the centre of the isophotal fitting to the nuclear region of M87 -assumed to mark the centre of mass of the galaxy- changes location depending on the image and size of the image analysed. In an absolute frame of reference, the change varies from 15 to 130 miliarcseconds (mas) with respect to the active galactic nucleus (AGN), which remains stable within an uncertainty of $\pm$15 mas in both x and y axis. The temporal analysis of the results indicates that the major displacements measured coincide with a powerful outburst that took place between 2003 and 2007, where there was a flux increment in the nucleus and the first knot of the jet. After the outburst, the isophotal centre remains stable and is consistent with the AGN location. This suggests that the displacements are artificially caused by a flux variation in the galaxy and that the SMBH actually resides in the equilibrium position. We caution about the determination of the galaxy photocentre by isophotal fitting in cases of nuclear variability and/or presence of photometric irregularities, and advise a long-term temporal analysis of the results to confirm possible displaced SMBHs.
\end{abstract}




\section{Introduction}
It is generally assumed that most massive galaxies contain a supermassive black hole (SMBH), which is expected to reside at the large-scale potential minimum of the host galaxy. However, there are several scenarios that would lead to a displacement of the SMBH from its equilibrium position \defcitealias{Batcheldor2010}{Paper~I}(\citealt{Batcheldor2010}, hereafter \citetalias{Batcheldor2010}). The possibilities include orbital motion of a SMBHs binary system, acceleration due to asymmetric radio jets, interaction with massive perturbers and gravitational recoil resulting from the coalescence of a SMBHs binary system. The latter mechanism has attracted the most interest in the past years. In this scenario, the merged SMBH can attain a kick velocity as a result of anisotropic emission of the gravitational waves \citep{Bekenstein1973,Favata2004}. Recent simulations of merging black holes predicted that recoil velocities of a few hundred to a few thousand km s$^{-1}$ can be reached \citep{Campanelli2007,Pretorius2007,Lousto2011,Blecha2016}. Such velocities would result in a variety of observational consequences, such as active galactic nuclei (AGNs) spatial and velocity offsets.\\
To date, several candidates have emerged from the search for recoiling SBHs \citep{Batcheldor2010,Civano2012,Lena2014,Kim2017,Skipper2018}. One of these candidates is the active giant elliptical M87 (NGC4486). This galaxy was studied by \citet{Batcheldor2010} and \citet{Lena2014}\defcitealias{Lena2014}{Paper~II} (hereafter, \citetalias{Lena2014}), who performed a photometric analysis of archival HST images with the aim of directly measuring a spatial offset between the AGN point source and the photocentre position. Their method assumes that the AGN marks the SMBH position and the photocentre, as defined by the galaxy isophotes, marks the minimum of the galactic potential. As a result, \citeauthor{Batcheldor2010} reported that the SMBH in M87 is relatively displaced to the galaxy photocentre position by $6.8 \pm 0.8$ pc in the counter jet direction, based on the analysis of two combined ACS images. They concluded that this displacement could be explained by jet acceleration and gravitational recoil. On the other hand, \citeauthor{Lena2014} analysed the same ACS images as \citetalias{Batcheldor2010} and four images taken with WFPC2 and NICMOS2. They found that the displacements measured in different images differ in amplitude and significance and they reported an error-weighted average displacement of $4.3 \pm 0.2$ pc roughly in the counter jet direction. Nevertheless, the origin of the discrepancies among \citetalias{Lena2014} results was unclear and no satisfactory explanation could be found. \\
Predictably, results relative to M87 have aroused great interest among the astrophysics community. This giant elliptical galaxy hosts one of the nearest and best studied AGNs. Furthermore, a relativistic jet is ejected from the galactic centre, where a SMBH of $(6.6 \pm 0.4) \cdot 10^{9} M_{\odot}$ \citep{Gebhardt2011} resides. The proximity of M87 has made its kiloparsec jet into a point of reference in the study of AGN jets. In addition, as M87 harbours one of the most massive SMBHs known, gravitational waves from the galaxy are expected to be detected with the Laser Interferometer Space Antenna (LISA). With all this, the study of the SMBH position is crucial to understand the evolution of this well known galaxy. \\
Here, we perform a photometric analysis of different HST and VLT images -some of the HST images used also by \citetalias{Batcheldor2010} and \citetalias{Lena2014}- with the aim of clarifying the presence of a photocentre-AGN displacement in M87. The method used is similar to that from \citetalias{Batcheldor2010} and \citetalias{Lena2014} and consists of measuring the relative distance between the AGN and the photocentre position. As a difference, we analyse images taken in different epochs and perform a temporal analysis of the results. Moreover, we add the size of the region analysed in the isophotal fitting as a new variable. These analysis allow us to obtain a more realistic view of the results found. We take the distance to M87 to be 16.7 Mpc, then 1\arcsec$\approx$81 pc  \citep{Blakeslee2009}.

\section{Data analysis}
\label{sec:data} 
The images analysed in this study and their characteristics are listed in Table \ref{table: images}. The data selected consists of twelve high spatial resolution images acquired with several instruments of the HST and the VLT. Optical and near-infrared (NIR) bands have been covered to control dust extinction if any.

\begin{table*}
\caption{Details of the images analysed.}
\begin{center}
\resizebox{1.0\textwidth}{!}{%
\begin{threeparttable}
\begin{tabular}{cccccccc}
\hline\hline
Name\tnote{a}   & Instrument & Filter - $\lambda_{c}$\tnote{b} & Pixel scale  & ID \tnote{c} & t \tnote{d} & Date & FWHM\tnote{e} \\
 & & &(arsec pixel$^{-1}$)& & (s) &  & (arsec)\\
\hline 

NaCo/K-06 \tnote{f}& VLT-UT4/NaCo & Ks - 2.18 $\mu m$ & 0.027 & 074.B-0404(A)    & 5 x 60 &23/01/2006& 0.15\\
NaCo/J-06 \tnote{f}&VLT-UT4/NaCo & J - 1.265 $\mu m$ & 0.027 & 074.B-0404(A)   & 8 x 98 & 23/01/2006& 0.24\\
ACS/HRC-03 \tnote{g,i} & HST/ACS/HRC & F814W - 833.3 $nm$ & 0.025 & 9829 & 4 x 50 &29/11/2003 - 07/02/2004 & 0.08\\
ACS/HRC-04 \tnote{g,i}& & & &10133&2 x 50&28/11/2004&0.09\\
ACS/HRC-05 \tnote{g,i}& & & &10133&50&09/05/2005&0.08\\
ACS/WFC-16 \tnote{i}& HST/ACS/WFC & F814W - 833.3 $nm$ & 0.050&13731&445&20/02/2016&0.12\\
WFC3-16 \tnote{i}& HST/WFC3/UVIS & F814W - 807 $nm$ & 0.040 & 14256 & 240&24/01/2016& 0.10\\
WFPC2-95 \tnote{i}&HST/WFPC2/PC & F814W - 801 $nm$& 0.046&5941&600&23/11/1995&0.10\\
WFPC2-98 \tnote{i} &  &  & &7274  &400&17/12/1998&0.10 \\
WFPC2-99 \tnote{i}& & & &8140&160&11/05/1999&0.11\\
WFPC2-01 \tnote{h,i}& & & &8592&260&25/06/2001&0.11\\
WFPC2-08 \tnote{i} & & &  & 11216&260&07/05/2008&0.10\\

\hline
\end{tabular}

\begin{tablenotes}
  \item[a] Name adopted in this work. Last digits indicate the observation year. \item[b] Filter and central wavelength. 
  \item[c] Number of identification of the observation proposal.
  \item[d] Exposure time (number of combined images x individual exposure time).
  \item[e] Full width at half maximum, computed in each image.
  \item[f] Images combined and reduced previously by the PARSEC group \citep{Montes2014}.
  \item[g] Images used by \citetalias{Batcheldor2010} and \citetalias{Lena2014} as part of a combined image.
  \item[h] Image also analysed by \citetalias{Lena2014}.
  \item{i} Images retrieved from the ESA Hubble Science Archive and with the standardpng STScI processing applied. 
  \end{tablenotes}
  \end{threeparttable}

\label{table: images}
}
\end{center}
\end{table*}

VLT adaptive-optics assisted NaCo images were combined and reduced previously by the PARSEC group \citep{Montes2014}. HST images were retrieved from the ESA Hubble Science Archive and had the standardpng STScI processing applied. However, ACS/HRC images are very noisy and we used a combination of images for ACS/HRC-03 and ACS/HRC-04. For these particular cases, each image was rotated and shifted to the same reference position before combining them with a median filter. Among the images analysed, ACS/HRC-03, ACS/HRC-04 and ACS/HRC-05 were used by \citetalias{Batcheldor2010} and \citetalias{Lena2014} as part of a combined image (see Table 1 in \citetalias{Batcheldor2010} for further details. Furthermore, WFPC2-01 was previously analysed in \citetalias{Lena2014} too. At the time of writing, ACS/WFC-16 and WFC3-16 are two of the most recent HST images available. In contrast to ACS/HRC, ACS/WFC and WFC3 have a wide field of view (202\arcsec x202\arcsec and 162\arcsec x162\arcsec respectively, versus 29\arcsec x25\arcsec from the ACS/HRC) that allows the analysis in a larger region of the images. All the images used in this work are shown in Appendix \ref{sec: appendix} (Fig. \ref{image1}, first column).

The method used in this study to determine the AGN-photocentre displacement is similar to that from the previous works \citetalias{Batcheldor2010} and \citetalias{Lena2014} and it is summarised as follows:
\begin{description}
 \item A mask was created to remove the irregularities such as globular clusters and the knots of the jet, which are visible in all the images. 
 \item An isophotal fitting to the galaxy surface-brightness distribution was performed within the central 9\arcsec radius of M87. The photocentre was computed as the average of the isophote centres.
 \item The position of the point nuclear source was determined by Gaussian distributions along the x and y axis.
 \item The displacement between the AGN and the isophotal centre and its position angle were computed.

\end{description}
The complete procedure is detailed in the next Sections \ref{sec: phot}, \ref{sec: AGN} and \ref{sec: dis}.

\subsubsection{Photocentre position}
\label{sec: phot}
\begin{figure*}
	\includegraphics[trim = 25mm 45mm 5mm 55mm, clip,width=1.0\textwidth]{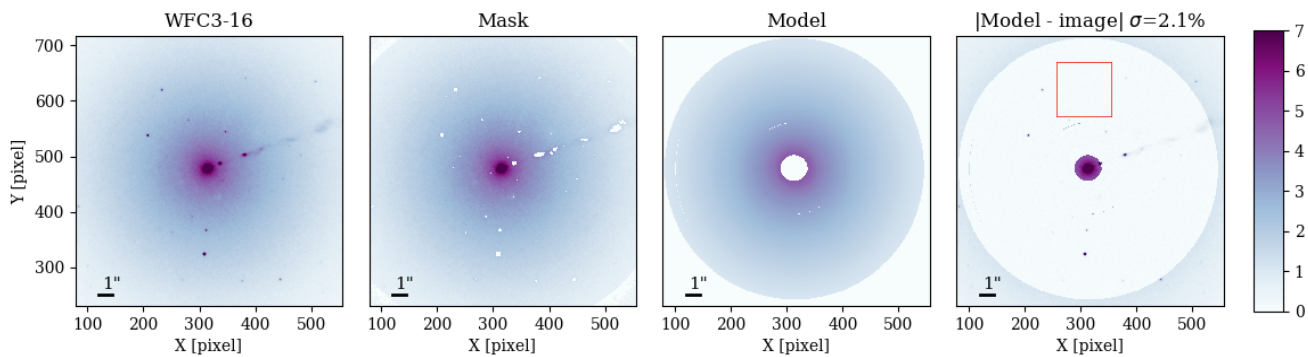}
    \caption{From left to right: original image obtained with WFC3- F814W, mask applied in the fitting, galaxy model computed from the fitted ellipses and galaxy after subtraction of the isophotal model. The percentage error between the original image and the model has been calculated in the square marked in the residual image.  }
    \label{fig: image}
\end{figure*}

The IRAF task \textit{ellipse} was used to fit elliptical isophotes to the data within a pre-specified region limited by a minimum and a maximum semi-major axis (SMA) value. Each image is measured using an iterative method described by \citet{Jedrzejewski1987}. This task produces an output of one table that contains the main characteristics for each individual isophote.\\
An important point in the procedure was to construct a mask for each image in order to minimise image defects and intrinsic photometric irregularities. 
Firstly, a run of \textit{ellipse} was performed and a galaxy model was created from the fitted isophotes. Secondly, the residual image was obtained and the mask was created by masking residuals exceeding the average signal. The central region of the galaxy was unmasked to allow \textit{ellipse} to find the isophotes centre. This method was iterated twice to get the final mask. Figure \ref{fig: image} shows an analysed image and its mask as an example. All the images analysed and their masks are included in Appendix \ref{sec: appendix} (Fig. \ref{image1}). Additionally, we have computed the percentage error $\sigma$ between the original image and its model. The percentage error has been estimated by dividing the median of the signal within a square region in the residual images and the median from the original images. The square regions used are plotted in the residual images in Fig.\ref{fig: image} and Fig. \ref{image1}. In all the cases the percentage error is fewer than 5$\%$, which indicates the reliability of the fitted models.\\
To determine the galactic photocentre, \textit{ellipse} was rerun with the final mask applied. According to \citetalias{Batcheldor2010}, isophotes within 1\arcsec are influenced by the AGN. Consequently, we adopted 1\arcsec as the minimum ellipse SMA. The maximum ellipse SMA was the one that fitted in the available field of view but never larger than the break radius of M87 nuclear region, $r= 9.41$\arcsec \citep{Capetti2005}. Between these limits the ellipse SMA was incremented by one pixel in each successive fit. The photocentre position was determined as the average of the individual ellipses centres. Increments of two pixels did not change the results.

\subsubsection{AGN position}
\label{sec: AGN}
This study follows on the premise that the point-like source at the centre of M87 coincides at all wavelengths, from UV to the IR, with the AGN location and the SMBH \citep{Prieto2016}. In this work, this position was determined by fitting a Gaussian function to the central point-like source of each image.

\subsubsection{Displacement and errors}
\label{sec: dis}
The displacements were computed as the difference between the isophote fitting centre and the AGN position, and its errors by the propagation of uncertainties. The task \textit{ellipse} returns the ellipse centre errors for each isophote fit, which vary from 0.1 to 0.6 pixels in our analysis. Nevertheless, there are larger deviations among the individual isophote centres, which can reach up to 3-4 pixels ($\la 0.15 \arcsec$). For that reason we adopted the error in the isophote fitting centre as the standard deviation, which shows the variability of the results obtained.

\section{Results}
\begin{figure*}
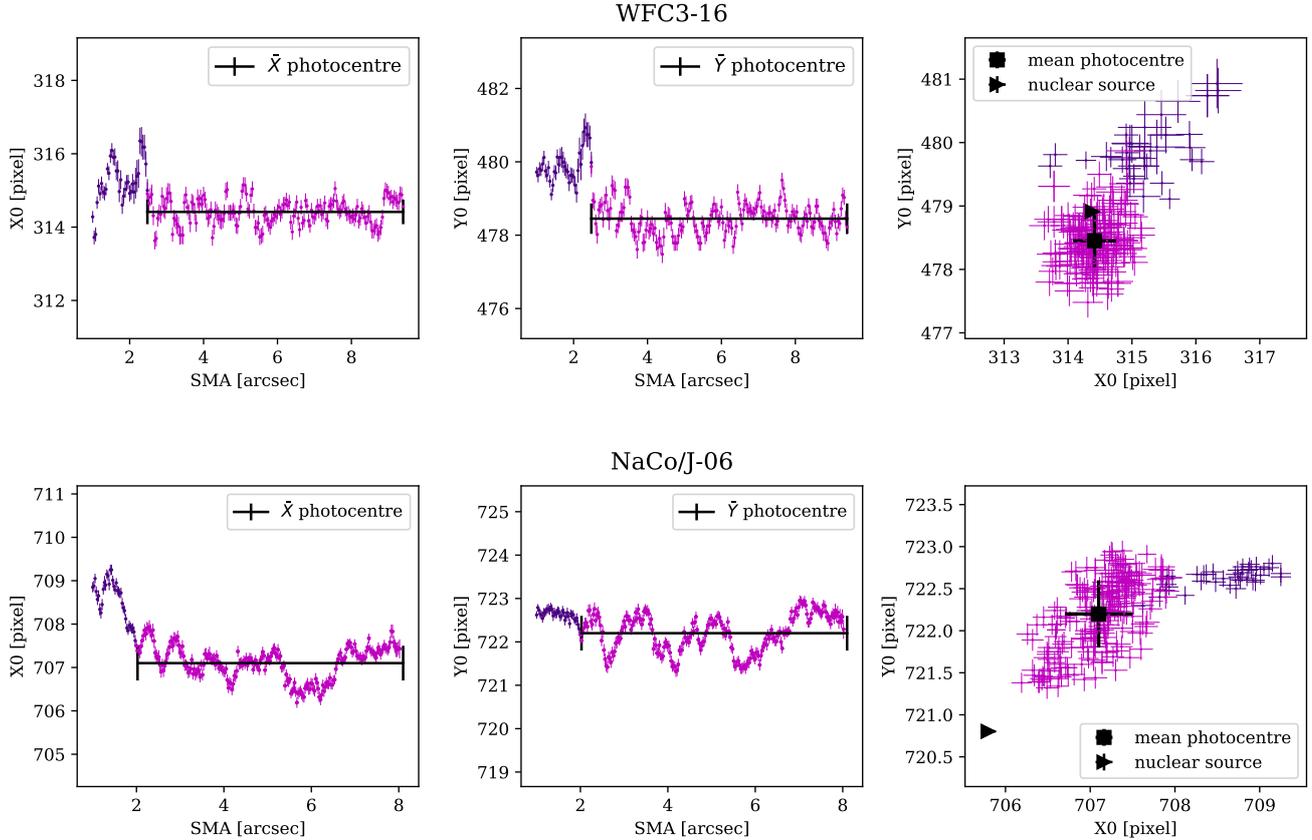

\subfigure{\includegraphics[trim = 25mm 84mm 25mm 10mm, clip,width=1.0\textwidth]{photoc_16}}\vspace{1mm}
\subfigure{\includegraphics[trim = 25mm 84mm 25mm 10mm, clip,width=1.0\textwidth]{photoc_06j}}
\caption{First and second column: X and Y coordinates of the individual isophote centres as a function of ellipse SMA. Third column: isophote centres and nuclear source positions. The mean isophotal fitting centre positions derived from $\approx$ 2\arcsec radius are included.  }
\label{fig: results}
\end{figure*}

Figure \ref{fig: results} illustrates two examples of the isophote fitting centre change as a function of the ellipse SMA (X and Y coordinates in the first and second column) and the isophotal centre coordinates compared to the AGN position (third column). Similar plots relative to all the images analysed are included in Appendix \ref{sec: appendix} (Fig. \ref{phot1}). It should be noted that in Fig. \ref{fig: results} there is an abrupt change in the isophotal centre at 2\arcsec approximately. For that reason we computed the mean isophotal centre in two different fitting regions, from 1 to 3\arcsec to be consistent with \citetalias{Batcheldor2010} and \citetalias{Lena2014}, and in a larger region from approximately 2\arcsec radius where the individual isophotal centres change softly. 

In Fig. \ref{fig: results} the mean isophote fitting centres derived from 2\arcsec radius are plotted, without considering the isophote centres that are closer to the AGN (plotted darker in the figure).  

The results of the photometric analysis are presented in Table \ref{table: results}. We list the ellipse SMA limits of the region analysed, the nuclear source and the isophotal fitting centre coordinates (x,y), the displacement between these positions and its position angle (in degrees E of N). The displacements in miliarcseconds units $\geq3\delta$ are highlighted in bold,  with $\delta$ being the displacement error.

{%
\newcommand{\mc}[3]{\multicolumn{#1}{#2}{#3}}
\begin{table*}
\begin{center}
\caption{ \textsc{Results of the images analysed.}}
\resizebox{1.0\textwidth}{!} {%
\begin{threeparttable}
\begin{tabular}{cccccccc}
\hline \hline
\textbf{Image} & \textbf{Nuclear source} ${X} \choose {Y}$ & \textbf{Region}\tnote{a} & \textbf{Isophotal centre} ${X} \choose {Y}$ & \mc{3}{c}{\textbf{Displacement $\pm \delta$}} & \textbf{Angle}\tnote{b}\\
 & (pixel) & (arsec) & (pixel) & (pixel) & (mas)& (pc)& (deg)\\\hline

NaCo/K-06& $699.29 \pm 0.04 $& [2:8] & $699.4 \pm 0.7$ & $2.1 \pm 0.6$ & \textbf{56$\pm$17}\tnote{c} & $4.6 \pm 1.4 $& $356 \pm 19$\\
 &  $714.01 \pm 0.03$ &  & $716.1 \pm 0.6$ &  &  &  &\\
 &  & [1:3] & $701.1 \pm 1.0$ & $2.4 \pm 0.8$ &  \textbf{64$\pm$21} &$5.1 \pm 1.7 $ &$311 \pm 18$\\
 &  &  & $715.6 \pm 0.4$ &  &  & & \\\hline

NaCo/J-06& $705.91 \pm 0.09 $& [2:8] & $707.1 \pm 0.4$ & $1.9 \pm 0.4$ &  \textbf{52$ \pm$11} &$4.2 \pm0.9 $ & $317 \pm 12$\\
 &  $720.803 \pm 0.004$ &  & $722.2 \pm 0.4$ &  &  & & \\
 &  & [1:3] & $708.0 \pm 0.7$ & $2.6 \pm 0.6$ & \textbf{71$\pm$16} & $6.0 \pm 1.3$ & $305 \pm 10$\\
 &  &  & $722.4 \pm 0.4$ &  &  & & \\\hline

ACS/HRC-03 & $190.73 \pm 0.03$ & [1:3] & $193.9 \pm 0.8$ & $4.0\pm 0.9 $ &  \textbf{100$\pm$21} &$8.1 \pm 1.8$ & $307 \pm 13$\\
 & $500.018 \pm 0.023$ & & $502.4 \pm 1.0 $ &  &  & & \\\hline
 
ACS/HRC-04 & $ 261.561\pm 0.008$ & [1:3] & $265.7 \pm 1.0$ & $5.2\pm 1.0 $ &  \textbf{130$\pm$24} &$10.6 \pm 2.0$ & $308 \pm 11$\\
 & $ 584.145\pm 0.026$ & & $587.4 \pm 0.9 $ &  &  & & \\\hline

ACS/HRC-05 & $277.603\pm 0.012$ & [2.5:5.1] & $278.2 \pm 1.0$ & $2.3\pm 1.1 $ &  $60\pm30$ &$4.7 \pm 2.3$ & $190 \pm 30$\\
 & $457.863 \pm 0.013$ & & $455.6 \pm 1.1 $ &  &  & & \\
 &  & [1:3] & $280.2 \pm 1.1$ & $2.6 \pm 1.2$ & $70\pm30$ & $5.3 \pm 2.4$ & $280 \pm 30$\\
 &  &  & $458.3 \pm 1.5$ &  &  & & \\\hline
 
ACS/WFC-16 & $ 1418.273\pm 0.015$ & [3:9.4] & $1417.94 \pm 0.17$ & $0.62\pm 0.21$ &  \textbf{31$\pm$10} &$2.5 \pm 0.8$ & $328 \pm 17$\\
 & $2049.817 \pm 0.015$ & & $2049.29 \pm 0.22 $ &  &  & & \\
  &  & [1:3] & $1418.7 \pm 0.4$ & $0.6 \pm 0.4$ & $28\pm21$ & $2.2 \pm 1.7$ & $280 \pm 30$\\
 &  &  & $2050.2 \pm 0.5$ &  &  & & \\\hline
WFC3-16 & $314.350 \pm 0.023$ & [2.5:9.4] & $314.4 \pm 0.3$ & $0.4 \pm 0.4 $ & $16 \pm 17$ &$1.3 \pm 1.3$ & $190 \pm 50$\\
 & $478.852 \pm 0.016$ & & $478.5 \pm 0.4 $ &  &  & &\\
 &  & [1:3] & $315.0 \pm 0.7$ & $0.9 \pm 0.6$ & $37 \pm 25$ & $3.0 \pm 2.0$ & $317 \pm 43$\\
 &  &  & $479.5 \pm 0.6$ &  &  & & \\\hline
 
WFPC2-95 & $ 163.10\pm 0.06$ & [2.5:6.8] & $165.4 \pm 0.4$ & $2.3 \pm0.4  $ &  \textbf{103$\pm$18} &$8.4\pm1.5 $ & $ 264\pm13 $\\
 & $279.27 \pm 0.05$ & & $ 279.0\pm 0.5 $ &  &  & &\\
 &  & [1:3] & $165.3 \pm 0.9$ & $2.6 \pm 0.8$ &  \textbf{120$\pm$40} & $10 \pm 3$ & $300 \pm 16$\\
 &  &  & $280.6 \pm 0.7$ &  &  & & \\\hline
WFPC2-98 & $157.298 \pm 0.003$ & [2.5:7.1] & $159.6 \pm 0.4$ & $2.3 \pm 0.4 $ &  \textbf{105$\pm$16} &$8.5 \pm 1.4$ & $271 \pm 11$\\
 & $301.66 \pm 0.03$ & & $301.7 \pm 0.4 $ &  &  & &\\
 &  & [1:3] & $159.7 \pm 0.9$ & $2.7 \pm 0.8$ &  \textbf{120$\pm$40} & $10 \pm 3$ & $295 \pm 15$\\
 &  &  & $302.8 \pm 0.6$ &  &  & & \\\hline
WFPC2-99 & $132.863 \pm 0.004$ & [2.5:6] & $133.2 \pm 0.5$ & $0.6 \pm 0.5 $ &  $28\pm22$ &$2.2 \pm 1.8$ & $220 \pm 50$\\
 & $327.25 \pm 0.05$ & & $326.8 \pm 0.4 $ &  &  & &\\
 &  & [1:3] & $133.4 \pm 0.4$ & $0.7 \pm 0.6$ & $30\pm30$ & $2.6 \pm 2.2$ & $310 \pm 50$\\
 &  &  & $327.7 \pm 0.7$ &  &  & & \\\hline
WFPC2-01 & $337.97 \pm 0.06$ & [2.5:9.4] & $338.6 \pm 0.4$ & $0.7 \pm 0.4 $ & $34 \pm 17$ &$2.7 \pm 1.4$ & $240 \pm 30$\\
 & $415.096 \pm 0.008$ & & $ 414.72\pm 0.4 $ &  &  & &\\
 &  & [1:3] & $338.8 \pm 0.5$ & $0.9 \pm 0.5$ & $42 \pm 24$ & $3.4 \pm 2.0$ & $290 \pm 40$\\
 &  &  & $415.4 \pm 0.6$ &  &  & & \\\hline

WFPC2-08 & $200.34\pm 0.06$ & [2.5:8.2] & $200.2\pm 0.3$ & $0.4 \pm 0.4 $ & $17 \pm 19$ &$1.3 \pm 1.6$ & $330 \pm 50$\\
 & $284.16\pm 0.04$ & & $283.8 \pm 0.4 $ &  &  & &\\
 &  & [1:3] & $201.0 \pm 0.6$ & $0.8 \pm 0.6$ & $36 \pm 29$ & $3.0 \pm 2.3$ & $300 \pm 40$\\
 &  &  & $ 284.5 \pm 0.6$ &  &  & & \\\hline

\end{tabular}
\begin{tablenotes}
  \item[a] Semi major axis limits where the isophotal fitting is performed. 
  \item[b] Isophotal centre position angle in degrees (E of N).
  \item[c] Displacements $\geq 3 \delta$ are highlighted in bold, with $\delta$ being the displacement error .
  \end{tablenotes}
  \end{threeparttable}
}
\label{table: results}
\end{center}
\end{table*}
}%

\section{Discussion}
The displacements between the AGN and the isophotal fitting centre found within a region from 1 to 3\arcsec radius are roughly aligned (within 30\degr) in the direction of the jet, which lies at P.A.$\approx$ 290$\degr$  \citep{Perez-Fournon1988}. This fact would imply that the SMBH is displaced from the equilibrium position in the counter-jet direction, as also concluded in \citetalias{Batcheldor2010} and \citetalias{Lena2014}. However, we notice that the displacements depend on the region of the galaxy analysed. If the isophotal fitting is performed from a 2\arcsec radius from the centre onwards, the displacements decrease in modulus and the position angle of the isophotal centre moves away from the jet direction. These discrepancies suggest that the isophotal fitting centre could be influenced by the flux that comes from the jet, specifically from the first knot HST-1 located at 0.85\arcsec from the AGN \citep{Perlman2003}. 

Results from NaCo, ACS-03, ACS-04, WFPC2-95 and WFPC2-98 images (Table \ref{table: images}) show significant displacements in both regions analysed, with a maximum amplitude of $\approx$0.1\arcsec. This result is consistent with that of the previous references \citetalias{Batcheldor2010} and \citetalias{Lena2014}. A detailed discussion of the possible displacement mechanisms was given by \citetalias{Batcheldor2010}, who concluded that the most likely scenarios that could explain the observed displacement were the jet acceleration and the gravitational recoil. \citetalias{Batcheldor2010} and \citetalias{Lena2014} used a combination of images from ACS obtained between 2002 and 2007, where M87 presented a high variability \citep{Perlman2003,Harris2006,Madrid2009}. In order to
reduce the margin of possible error as much as possible, we analysed ACS images separating them in epochs. From these images we recovered \citetalias{Batcheldor2010} and \citetalias{Lena2014} ACS results in ACS-03 and ACS-04 images (Table \ref{table: images}).

On the other hand, the photocentre position derived from WFC3-16, WFPC2-99, WFPC2-01 and WFPC2-08 images (Table \ref{table: images}) are consistent with the AGN position. In order to find a possible explanation to the discrepancies of the results derived from different images we studied the influence of the mask used in the analysis. We found that the fit is very sensitive to photometric irregularities such as M87 globular clusters. Moreover, in the calculation of the first isophotes the mask was crucial to hide HST-1, that significantly affects the isophotal fitting centre in the region from 1 to 3\arcsec radius. 

It is remarkable that the discrepancies in the displacement amplitude among the images analysed are similar to that from \citetalias{Lena2014}, with the main difference that we have used different data.

\subsection{Comparison in an absolute frame of reference}

\begin{figure}
	\subfigure{\includegraphics[width=1.0\columnwidth]{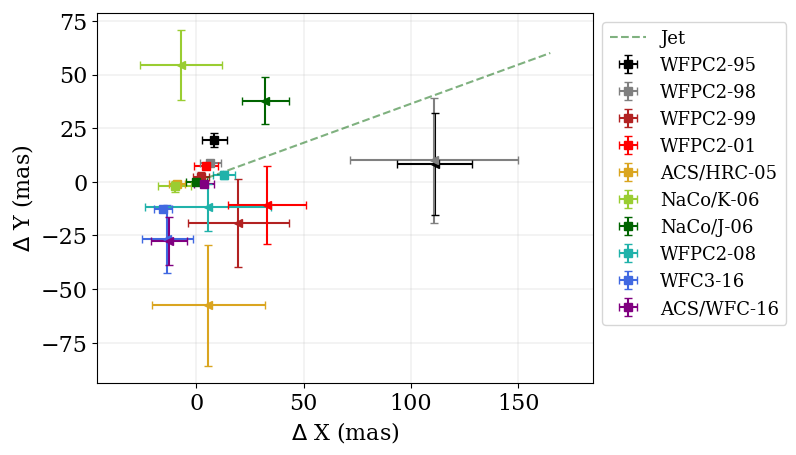}}
	\subfigure{\includegraphics[width=1.0\columnwidth]{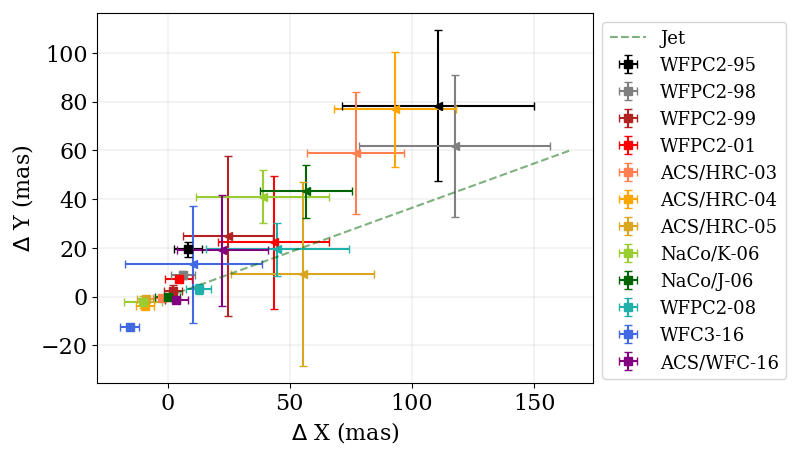}}
    \caption{Nuclear sources (squares) and isophote fitting centres (triangles, with larger errors) obtained from the images analysed. Top: isophote centres derived from $\approx$2\arcsec radius. Bottom: isophote centres computed within 1 and 3\arcsec. The jet mean direction is marked with a discontinuous line. All positions are plotted respect to an absolute frame of reference defined by the mean position of 5 globular clusters in M87. The origin is centred to the NaCo-J image nuclear source.      }
    \label{fig: absframe}
\end{figure}

\citetalias{Lena2014} performed a comparative analysis to determine if their results were consistent with a change in the AGN or in the isophotal fitting centre. They found that the differences in the displacements were due largely to differences in the isophotal fitting centre, but the origin of these discrepancies was unclear. Similarly, we compared the results from all the images analysed in this work defining an absolute frame of reference based on the position of five globular clusters in the central 12$\arcsec$x12$\arcsec$ of M87, which are assumed stable over time. Their positions were determined by fitting Gaussians to their light distributions profiles with the IRAF task \textit{center}. The absolute frame of reference was determined as the average position of the five globular clusters. Finally, the positions relative to this frame of reference were computed and transformed to physical units (mas). Figure \ref{fig: absframe} shows the results of the comparison. The isophotal fitting centres obtained in the M87 region beyond 2\arcsec radius and in between 1 and 3\arcsec are plotted in the upper and lower graph respectively. We centred the origin in the AGN position of the NaCo-J image to make the comparison easier. The uncertainties associated to the change of the frame of reference were propagated and added to the individual position errors. The jet mean direction from NaCo-J image is marked with a discontinuous line.

First of all, we observe that the AGN remains fairly stable within an error of $\pm$15 mas in both the x and y directions. Nevertheless, we noticed that there is a tendency in the AGN positions to align in the jet direction. This alignment could be due to the emission from the most inner knots of the jet, which are not spatially resolved in the images analysed \citep{Hada2016}.\\
On the contrary, the isophotal fitting centres vary depending on the image and on the region analysed within the image. This fact implies that the inferred isophote centre cannot be identified with M87 photocentre and therefore with the large-scale potential minimum of M87 as it is assumed in this method. Any variation in the equilibrium configuration of such massive galaxy would require a much larger time scale than the time period of 21 years probed in this analysis. Moreover, the isophotal centres derived within the region from 1 to 3\arcsec are clearly aligned along the jet, whereas these positions computed from 2\arcsec do not present any obvious tendency. This suggests that the isophote fitting centre change may be related to a variation of the flux that comes from the AGN or the jet, even though both contributions are masked in the isophotal analysis.  

\subsection{Temporal change}
\begin{figure}
\includegraphics[trim= 15mm 2mm 28mm 20mm, clip,width=1.0\columnwidth]{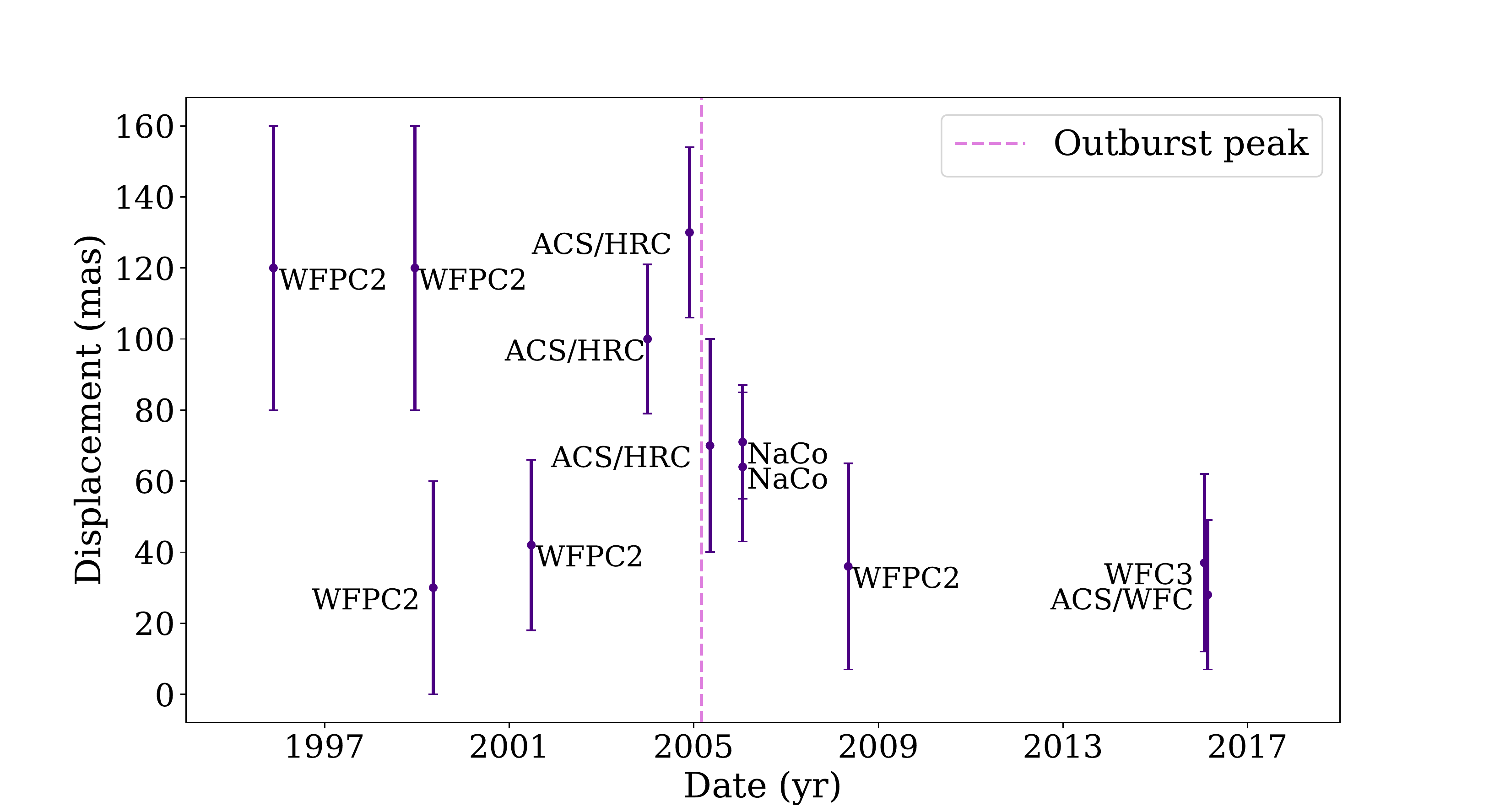}
	
    \caption{Displacements of the isophote fitting centre derived from 1 to 3\arcsec radius versus the observation dates of the images analysed. HST images were taken in the F814W filter. VLT/NaCo images were obtained in the J and K bands.   }
    \label{fig: variation}
\end{figure}

 In an attempt to understand the isophotal fitting centre change we performed a temporal analysis of the results. We noticed that most images that exhibit a displaced isophote centre were obtained during the same period of time where there was a powerful outburst in M87. This event occurred between 2003 and 2007 and led to a variation of the output of HST-1 and the nuclear source in all wavebands. There was a reported increase of HST-1 intensity by a factor of 50 in X rays between 2000 and 2005 \citep{Harris2006} and a factor of 90 in NUV \citep{Madrid2009}, becoming brighter than the nucleus of the galaxy. This event was accompanied by fast variations of the tera-electron volt (10$^{12}$ eV) gamma-ray flux in scales of days, with an emission peak between February and May 2005 \citep{Aharonian2006}.\\
 Figure \ref{fig: variation} shows the temporal change of the isophotal centre displacements derived from our analysis. The isophotal centres obtained between 1 and 3\arcsec are plotted, since it is the region where a larger influence of the outburst is expected. \\

 According to this analysis, the variation of the isophote centre is clearly related to the flux increase of HST-1. The maximum displacement found in this study took place at the end of 2004, close to the emission peak of the outburst. After the event, our results indicate a stability of the isophote centre in a position consistent with the AGN. Consequently it is reasonable to consider that the AGN actually remains in the potential minimum of the galaxy, and therefore the SMBH is not displaced from its equilibrium position. \\
 Surprisingly, we analysed two images obtained before the outburst (WFPC2-95 and WFPC2-98) that exhibit a significant displacement that cannot be explained with the flux variability in HST-1, as in that epoch this knot was almost imperceptible. The origin of these displacements is unclear. We performed a preliminary analysis of the nuclear source flux and we found no evidence of any anomalies during those years. Multiple epochs VLBA observations of M87 show that the flux density and brightness temperature of the core vary relatively smoothly over time and that the location of five separate features within the inner 20 mas from the nucleus appeared essentially stationary from 1995 to 2007 \citep{Kovalev2007}. Therefore, we can relate the displacements found in WFPC2-95 and WFPC2-98 images neither to a flux variation from the nuclear source nor from HST-1. Since these displacements are roughly aligned along the jet axis and taking into account M87 intrinsic variability, we guess that other possible explanation could be the appearance of a new blob of plasma along the jet during those years. In this case, the flux density would influence the isophotal fitting in a similar way as the outburst. Facing the impossibility of analysing previous images, this explanation is only speculation. For lack of variability reports before the outburst, it might be also thought that the photocentre location found in WFPC2-95 and WFPC2-98 images is the actual location. However, as what varies more significantly during the outburst is HST-1 (which is also along the jet direction), its flux density should cause a displacement added to that from WFPC2-95 and WFPC2-98, which is not the case. Therefore we believe that these displacements are also artificial and that the photocentre position derived from WFPC2-95 and WFPC2-98 do not indicate the potential minimum of the galaxy.  \\
 It is important to emphasise that according to this temporal analysis the fast variation of the isophote fitting centre is due to an additional contribution to the habitual flux of the galaxy. This is a satisfactory explanation for the differences in the displacement position angle and magnitude. In the images analysed during the outburst, the additional flux comes from the first knot of the jet HST-1, which becomes even brighter than the nucleus. Therefore, the closer to HST-1 we analyse the image, the smaller the elliptical isophotes are and the more the additional flux influences them.
 
 Changes in the photocentre position of different AGNs have been reported by other authors \citep{Anton2012,Popovic2012}. \citet{Anton2012} found that highly variable quasars show photocentre variations about tens of parsec that are accompanied by fluctuations in the output energy. They discussed that the most likely origins of this instability in radio-louds AGNs are enhancements in flux due to shocks along the jet or the appearance of new radio blobs of plasma. 
However, \citeauthor{Anton2012} assumed that the photocentre is actually the nuclear source. Therefore, the similarities with our results must be taken carefully: if variability in the nuclear source is able to displace the centroid, then the isophotes centres could be affected by a flux variation too.
 
\begin{figure}
 \includegraphics[width=\columnwidth]{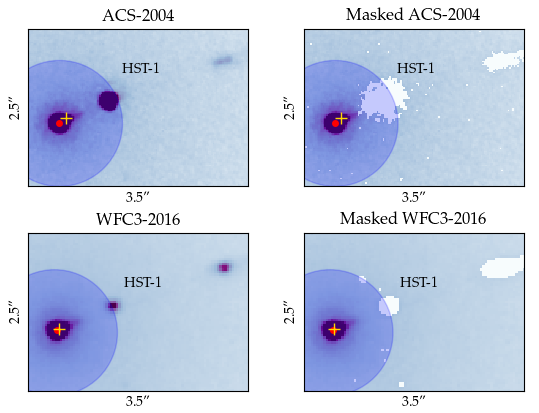}
	
    \caption{Central regions of two images analysed. First column: original images. Second column: masked images. The nuclear sources (points) and the isophotal centres (crosses) derived within the region [1:3]\arcsec are plotted in each image. The nuclear source is at the coordinate (0.5,1) and the circumference shows the region not analysed in the fitting (until 1\arcsec radius).}
    \label{fig: perspective}
\end{figure}

 With the aim of getting a more realistic understanding of the results, a central region of some original and masked images are presented in Fig. \ref{fig: perspective}. The nuclear source and the isophote centre derived within the region from 1 to 3\arcsec radius are plotted in each image. On the basis of the great intensity of the outburst, the displacement dimensions compared to the nuclear source size favour the idea that actually HST-1 could influence the isophotal centre even masking the knot itself. Moreover, the accuracy required in this method also suggests that the displaced isophote centres found from 2\arcsec radius are due to a scattering of data. A statistical study with more images would be necessary to confirm this theory and therefore this point cannot be concluded in this paper.
 
\section{Conclusions}
The aim of this study is to clarify the existence of a displaced SMBH in M87, which was argued to be the case in previous works in the literature. We have determined the photocentre position of several HST and VLT adaptive optics images taken at different epochs by means of an isophotal fitting of the central 9\arcsec radius of M87. The photocentre- AGN displacement was estimated assuming that the SMBH is at the AGN location, the latter being identified as a point-like source from UV to IR bands. From the analysis of the results we can draw the following points:
\begin{enumerate}
\item Displacements between the isophotal fitting centre and the AGN location from different images differ in amplitude and position angle. From the same set of images used in previous works \citepalias{Batcheldor2010,Lena2014} our values are consistent with those found by these authors. The maximum displacement found is $\approx$0.1\arcsec (8.1 pc). However, in some images the isophotal centre is consistent with the AGN location. The calculation of the positions in an absolute frame of reference, which is defined by the average position of five globular clusters in the central 12$\arcsec$x12$\arcsec$ of M87, indicates that the origin of the displacements when found is due to an isophotal fitting centre change and that the AGN remains stable within an uncertainty of $\pm$15 mas in both x and y directions. 

\item The isophotal fitting centre depends on the region analysed in each image. When the isophotal fitting is restricted to an annulus between 1 and 3\arcsec radius from the centre, the displacements are roughly aligned in the jet direction (PA$\approx$290$\degr$). If the isophotal fitting starts instead from $\approx$ 2\arcsec onwards, the displacements decrease and move away from the jet direction. The temporal analysis of the results exhibits a link between the displacements and a powerful outburst that took place from 2003 to 2007 in M87. This event led to an increase of the flux density in the nuclear source and in the first knot of the jet, HST-1, during those years. This flux increase explains the discrepancies of the results from different images and computed in separated regions, i.e. the flux increase correlates with the displacements of the isophotal centres.    

\item After M87 outburst, the apparent stability of the isophotal fitting centre in a position that is consistent with the AGN favours the idea that the SMBH is actually in the potential minimum of M87. However, we found two images obtained before the outburst that exhibit a significant displacement. The origin of these results is unclear. We speculate that events related to the intrinsic variability of M87 could have caused these displacements.

\end{enumerate}
We conclude that there is strong evidence that the presence of a displacement between the SMBH and the photocentre in M87 is due to an isophotal fitting centre change caused by M87 nuclear-jet flux variation, and not to displacements of the SMBH as interpreted in previous works \citepalias{Batcheldor2010,Lena2014}. All things considered, we determine that the SMBH location in M87 remains at the photocentre of the galaxy.

We caution about the photocentre determination in cases of nuclear variability and/or the presence of additional sources close to the nucleus of a galaxy, like the knots of a jet, and advise a long-term temporal analysis of the data to confirm possible displaced SMBHs or a SBMHs binary in the centre of galaxies.

Gaia space mission will find possible candidates of BH motions in centre of quasars. However, high angular resolution will still be needed together with the temporal analysis to disentangle the cause of the nuclear motion, as the presence of emitting blobs in optical jets could be detected as proper motion shifts. 
\section*{Acknowledgements}
We would like to thank Dr. J. A. Fern\'andez-Ontiveros and the members of PARSEC group for all the comments and discussions. We thanks Dr. M. Schartmann and Dr. D. Calderon  for comments on the manuscript.




\bibliographystyle{mnras}
\bibliography{paper}



\appendix
\section{Images and results}
\label{sec: appendix}

\begin{figure*}
\centering

\subfigure{\includegraphics[trim = 25mm 45mm 5mm 55mm, clip,width=1\textwidth]{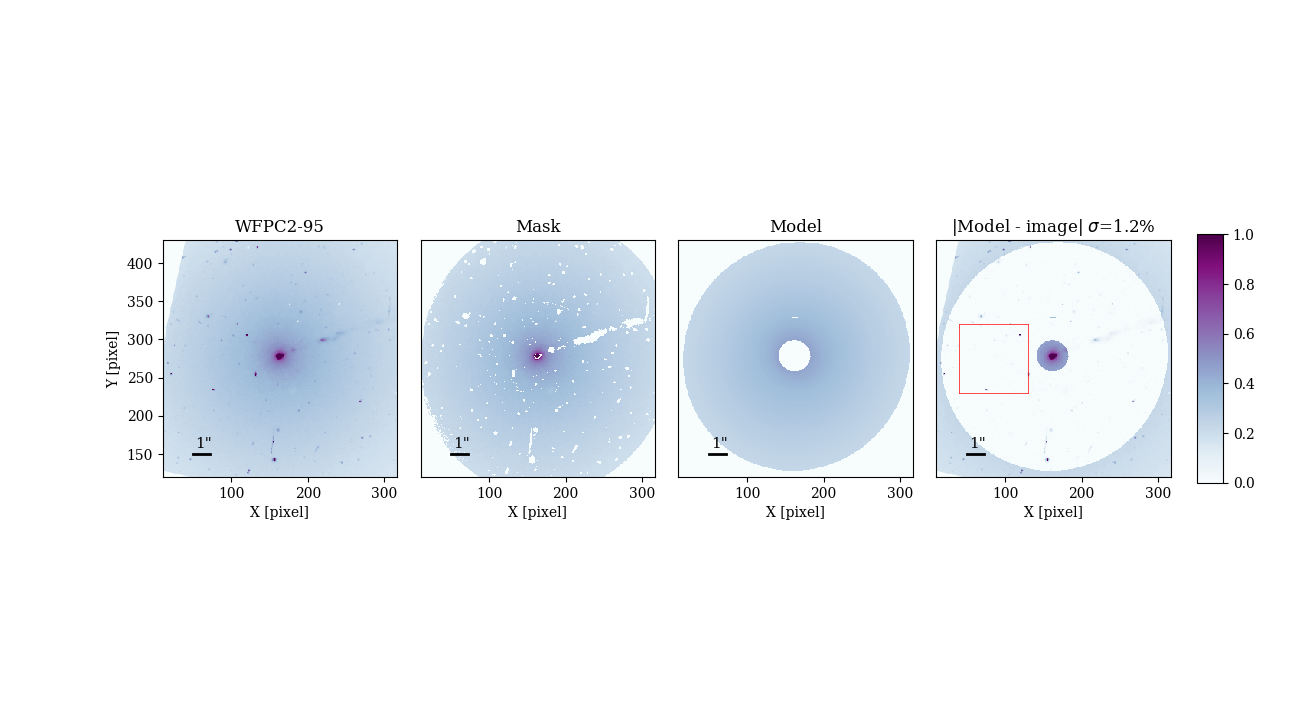}}\vspace{0mm}
\subfigure{\includegraphics[trim = 25mm 45mm 5mm 55mm, clip,width=1\textwidth]{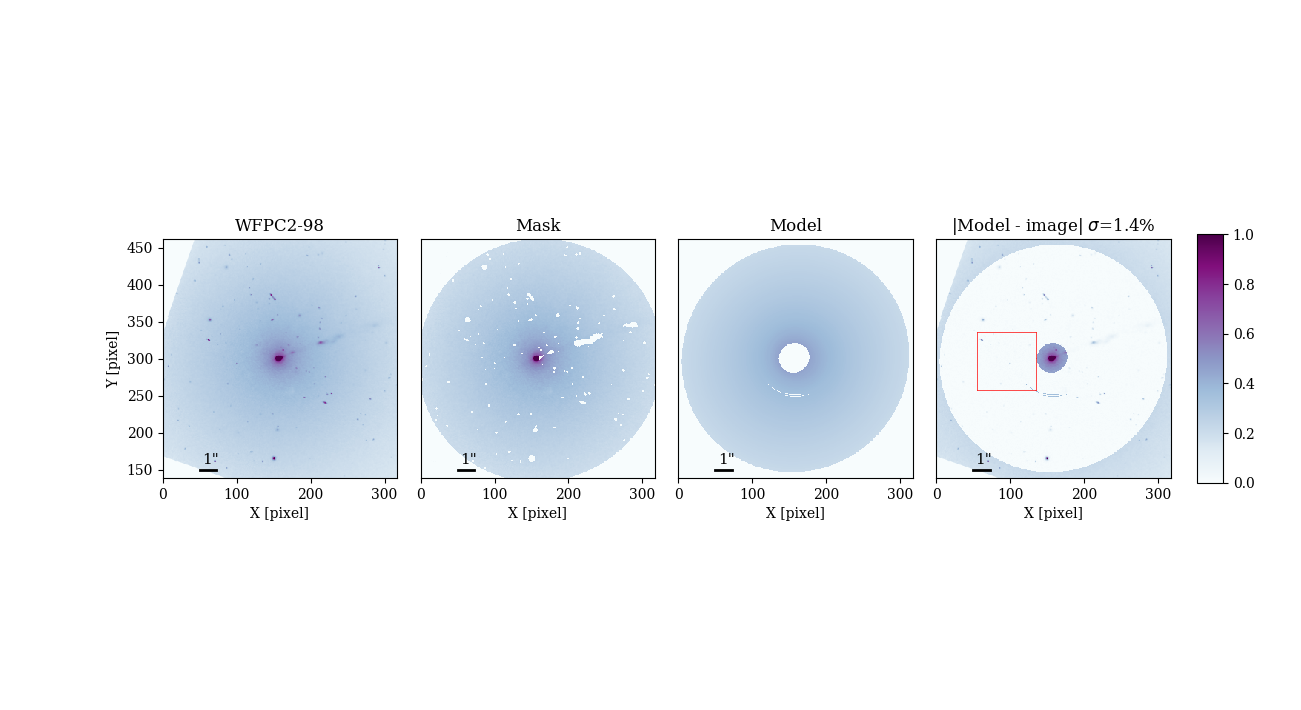}}\vspace{0mm}
\subfigure{\includegraphics[trim = 25mm 45mm 5mm 55mm, clip,width=1\textwidth]{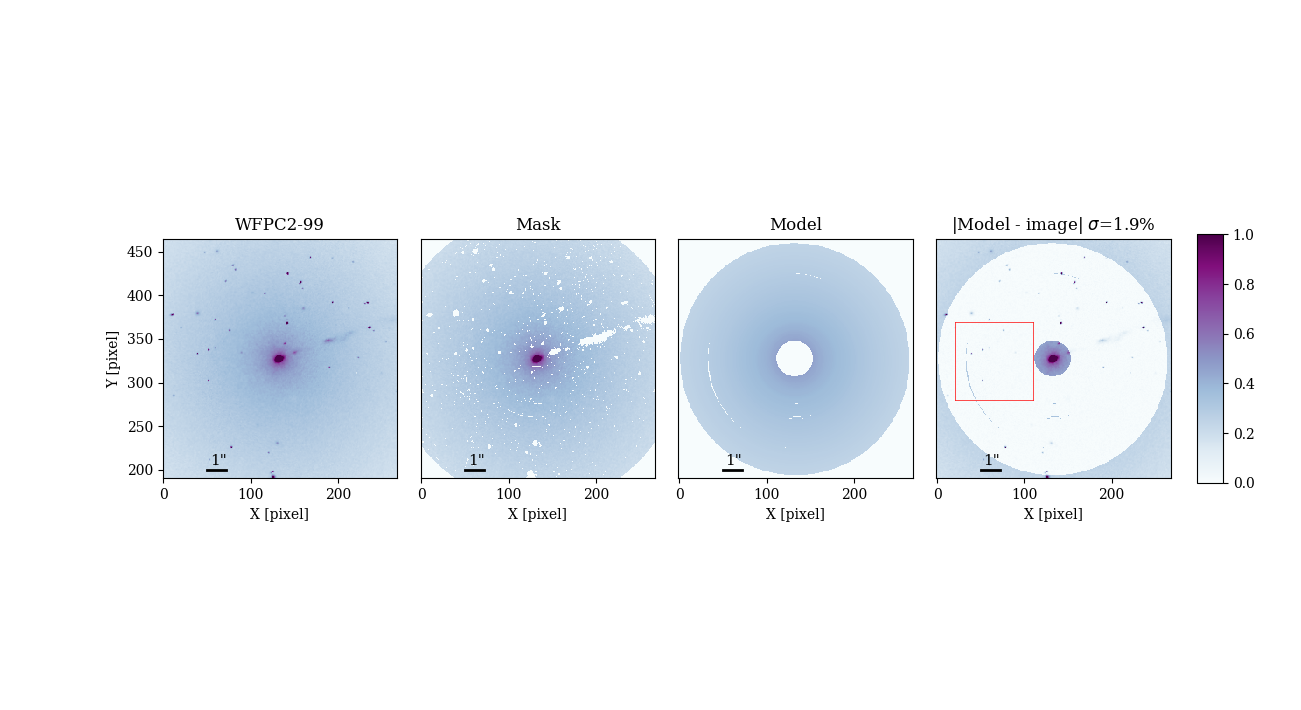}}\vspace{0mm}
\subfigure{\includegraphics[trim = 25mm 45mm 5mm 55mm, clip,width=1\textwidth]{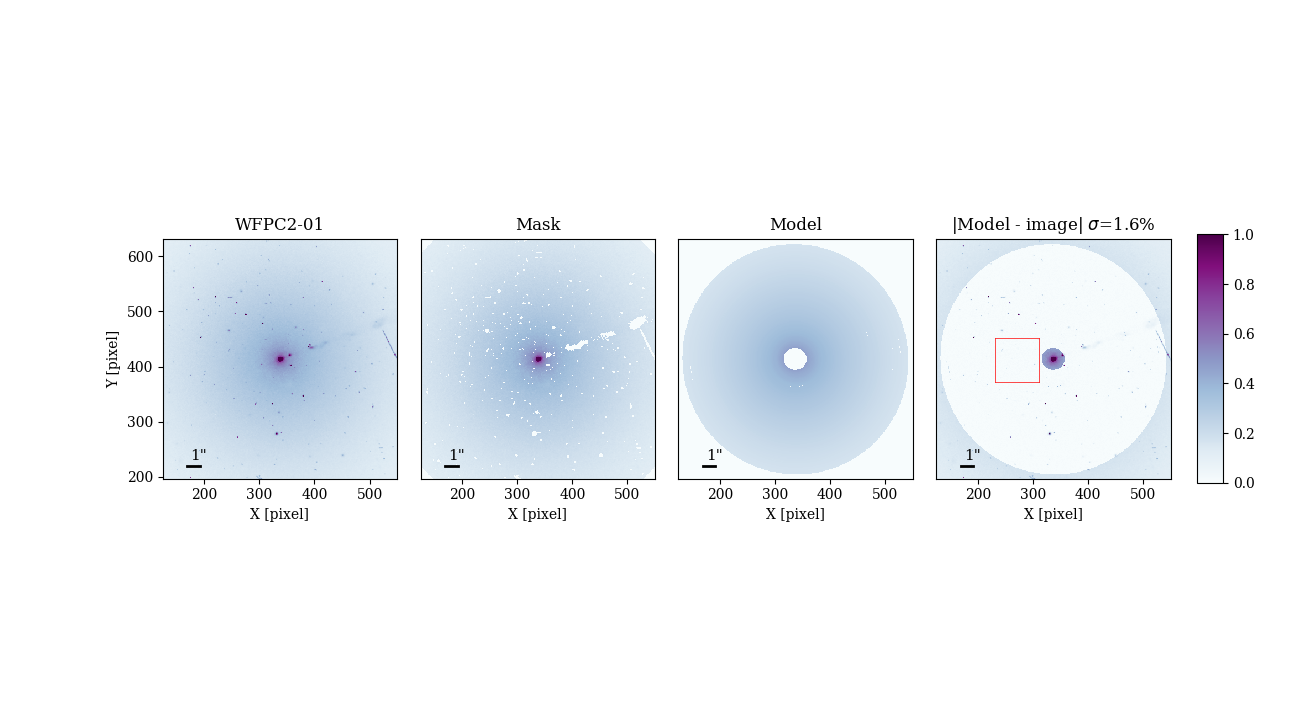}}\vspace{0mm}

\caption{From left to right: original image, mask applied in the fitting, galaxy model computed from the fitted ellipses and galaxy after subtraction of the isophotal model. The percentage error between the original image and the model has been calculated in the square marked in the residual image.   }
\label{image1}
\end{figure*}
 
 \begin{figure*}
\centering
\subfigure{\includegraphics[trim = 25mm 45mm 5mm 55mm, clip,width=1\textwidth]{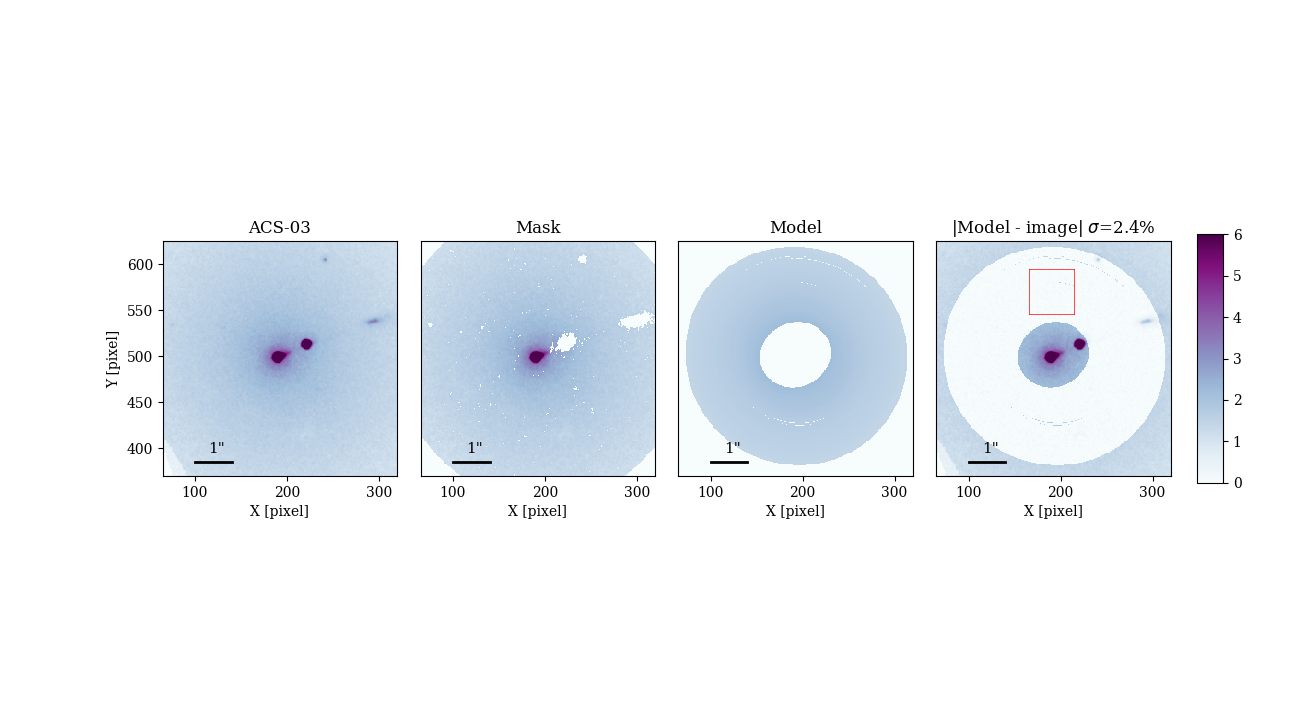}}\vspace{0mm}
\subfigure{\includegraphics[trim = 25mm 45mm 5mm 55mm, clip,width=1\textwidth]{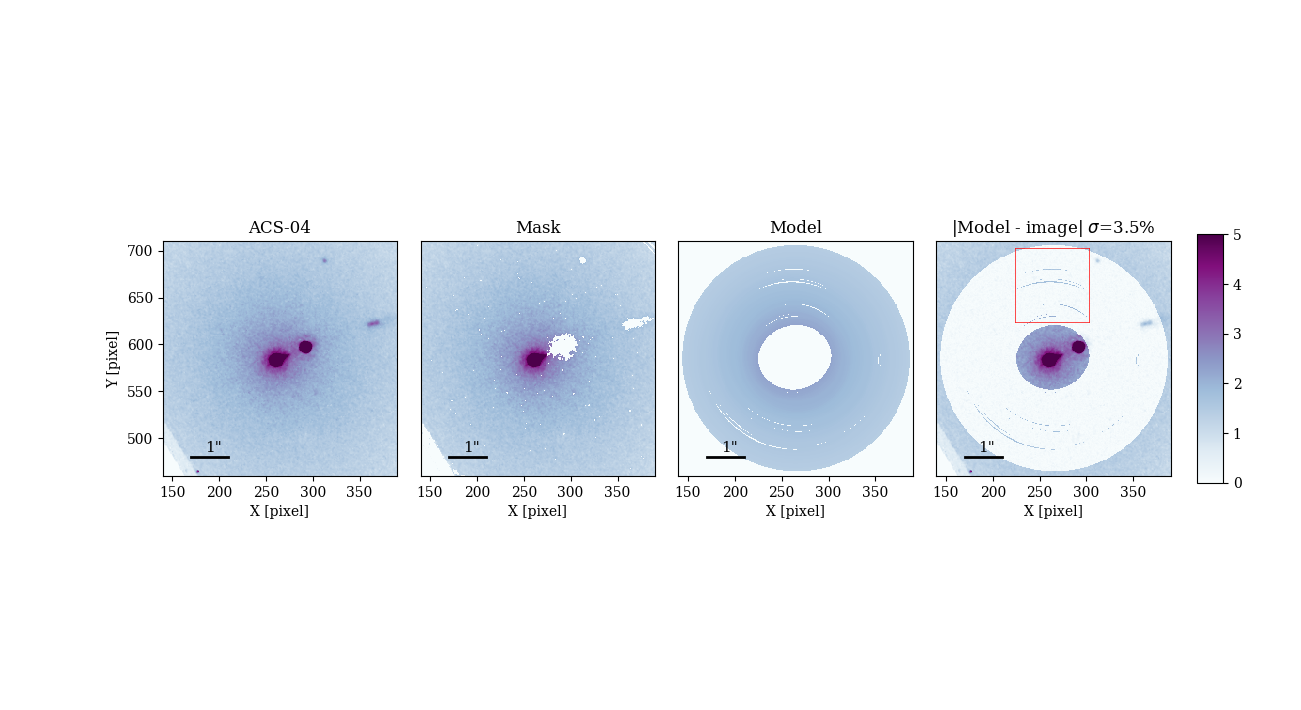}}\vspace{0mm}
\subfigure{\includegraphics[trim = 25mm 45mm 5mm 55mm, clip,width=1\textwidth]{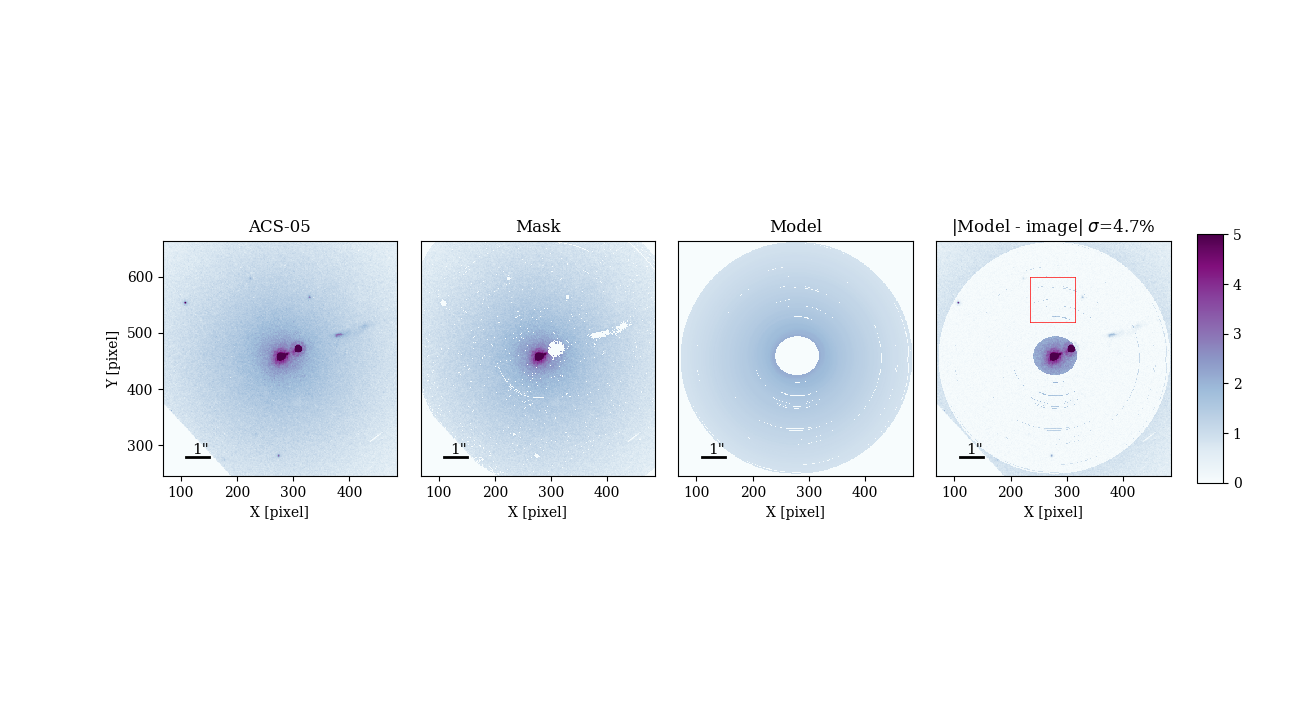}}\vspace{0mm}
\subfigure{\includegraphics[trim = 25mm 45mm 5mm 55mm, clip,width=1\textwidth]{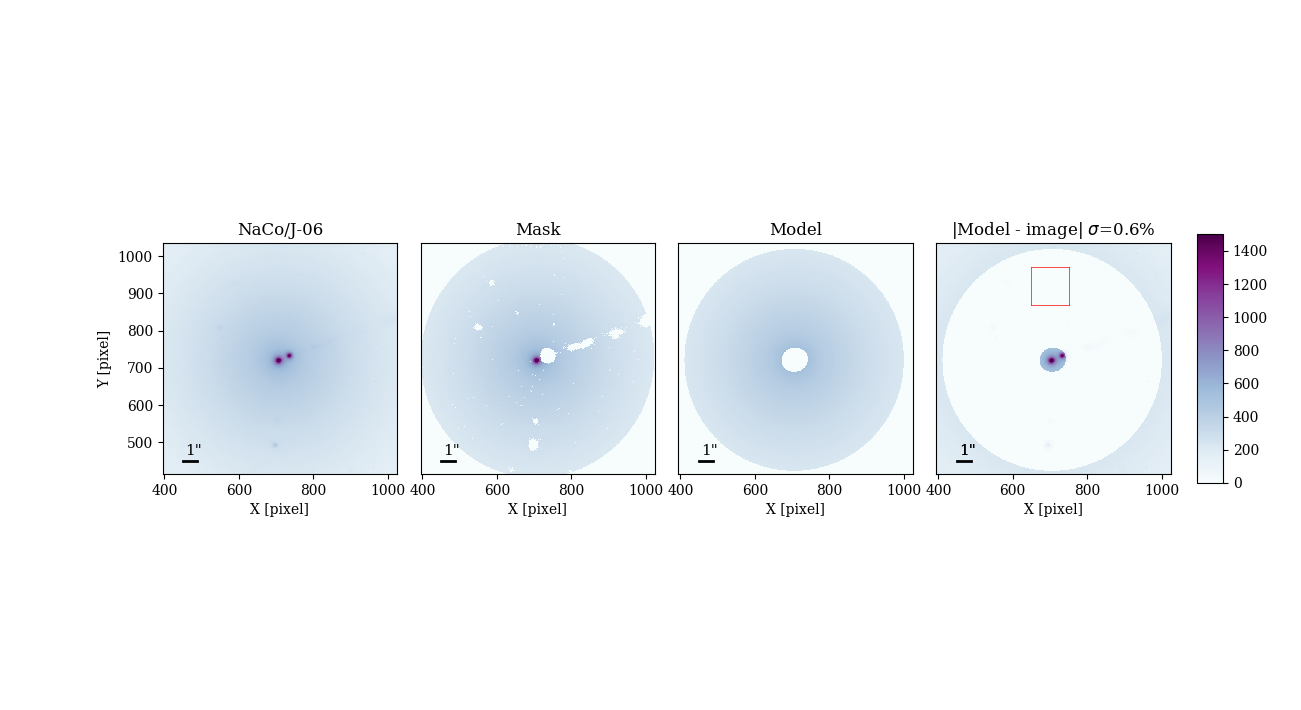}}\vspace{0mm}

\contcaption{From left to right: original image, mask applied in the fitting, galaxy model computed from the fitted ellipses and galaxy after subtraction of the isophotal model. The percentage error between the original image and the model has been calculated in the square marked in the residual image.   }
\label{image2}
\end{figure*}

 \begin{figure*}
\centering
\subfigure{\includegraphics[trim = 25mm 45mm 5mm 55mm, clip,width=1\textwidth]{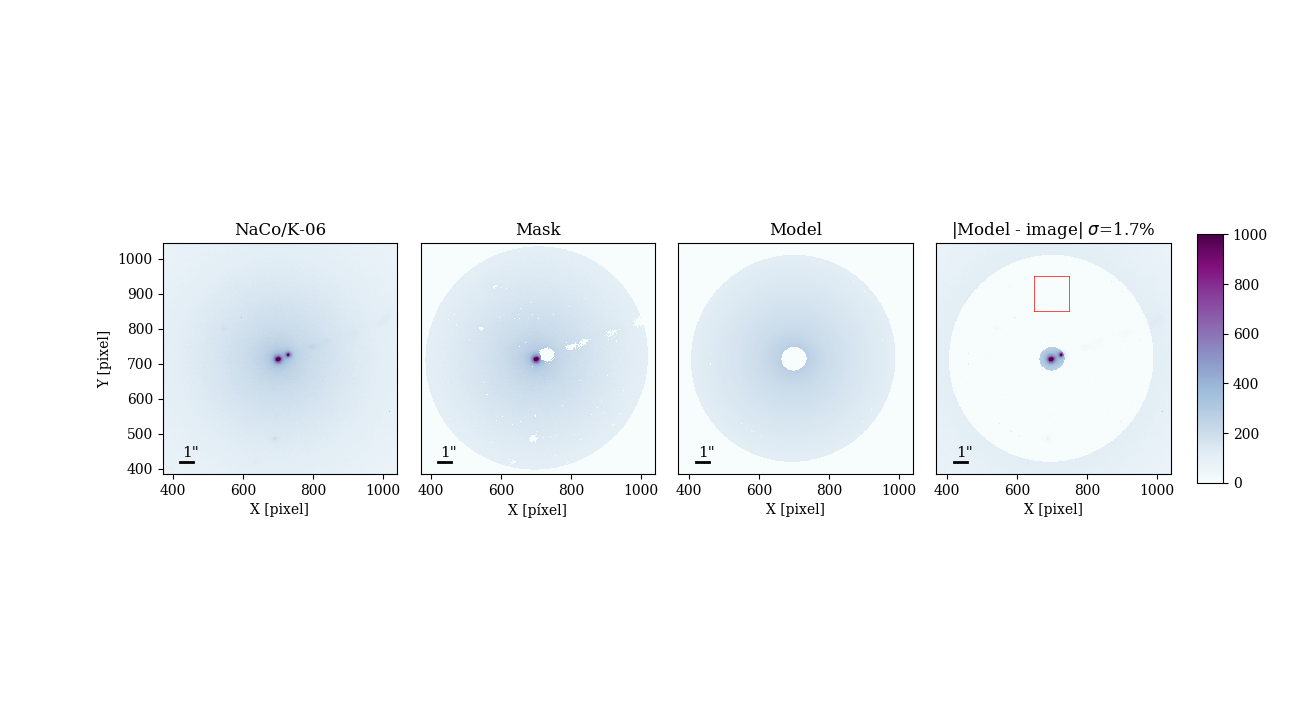}}\vspace{0mm}
\subfigure{\includegraphics[trim = 25mm 45mm 5mm 55mm, clip,width=1\textwidth]{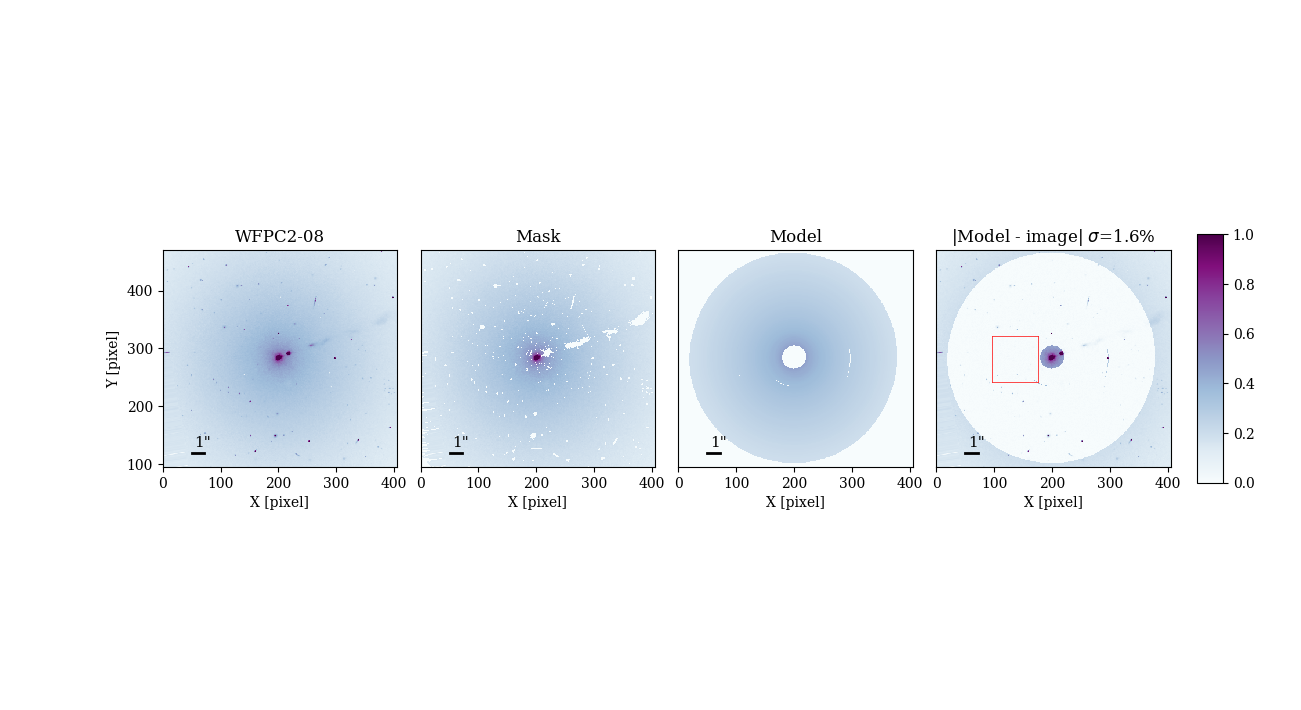}}\vspace{0mm}
\subfigure{\includegraphics[trim = 25mm 45mm 5mm 55mm, clip,width=1\textwidth]{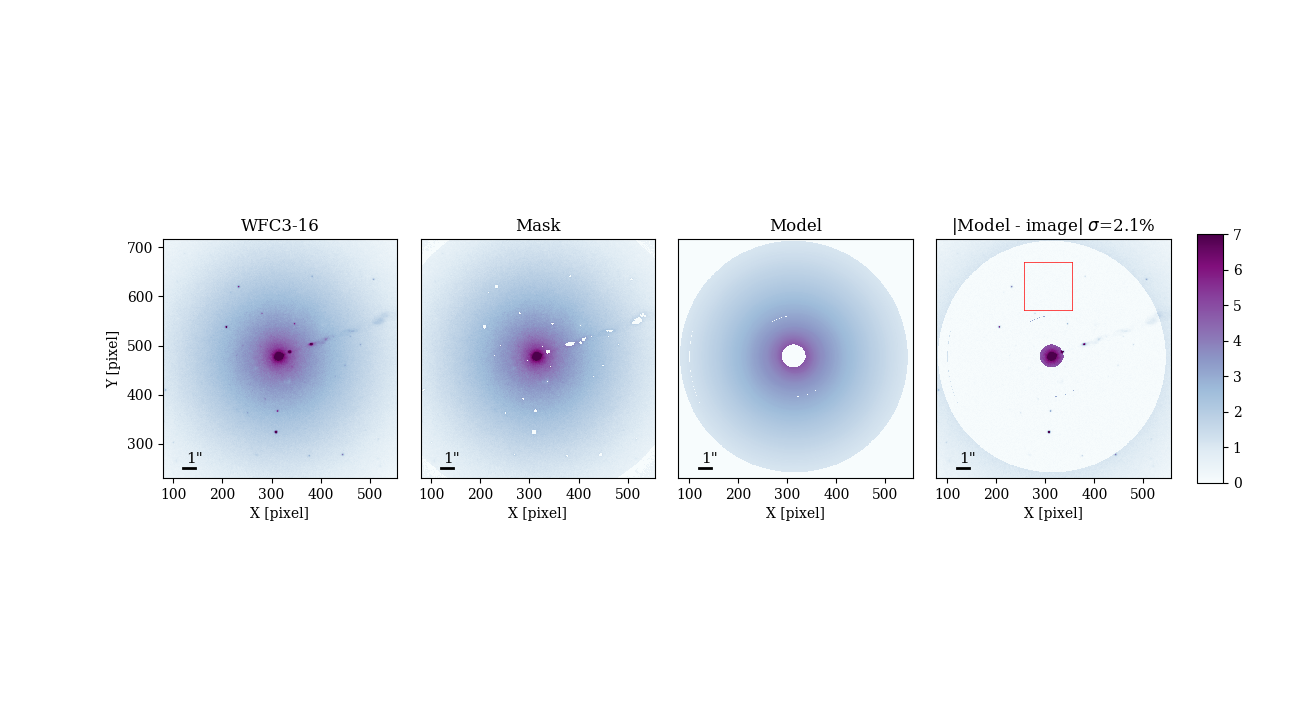}}\vspace{0mm}
\subfigure{\includegraphics[trim = 25mm 45mm 5mm 55mm, clip,width=1\textwidth]{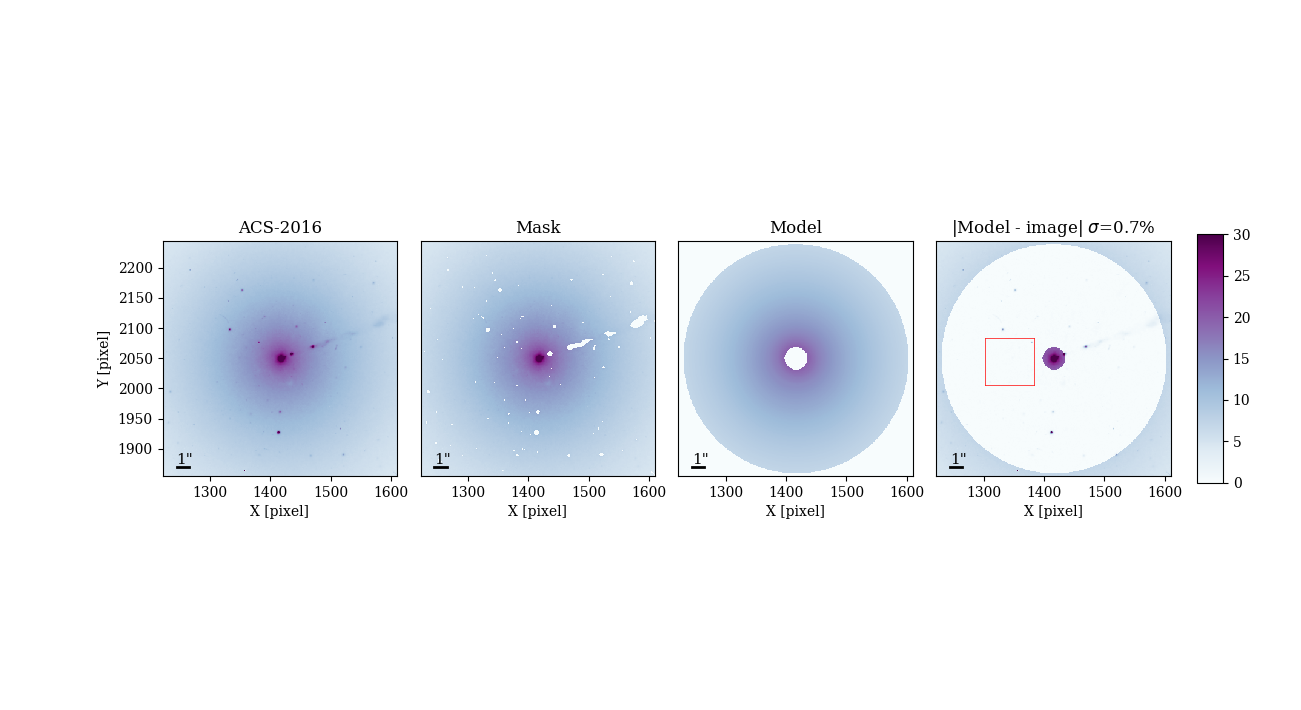}}\vspace{0mm}
\contcaption{From left to right: original image, mask applied in the fitting, galaxy model computed from the fitted ellipses and galaxy after subtraction of the isophotal model. The percentage error between the original image and the model has been calculated in the square marked in the residual image.   }
\label{image3}
\end{figure*}

\begin{figure*}
\centering

\subfigure{\includegraphics[trim = 25mm 84mm 25mm 10mm, clip,width=1\textwidth]{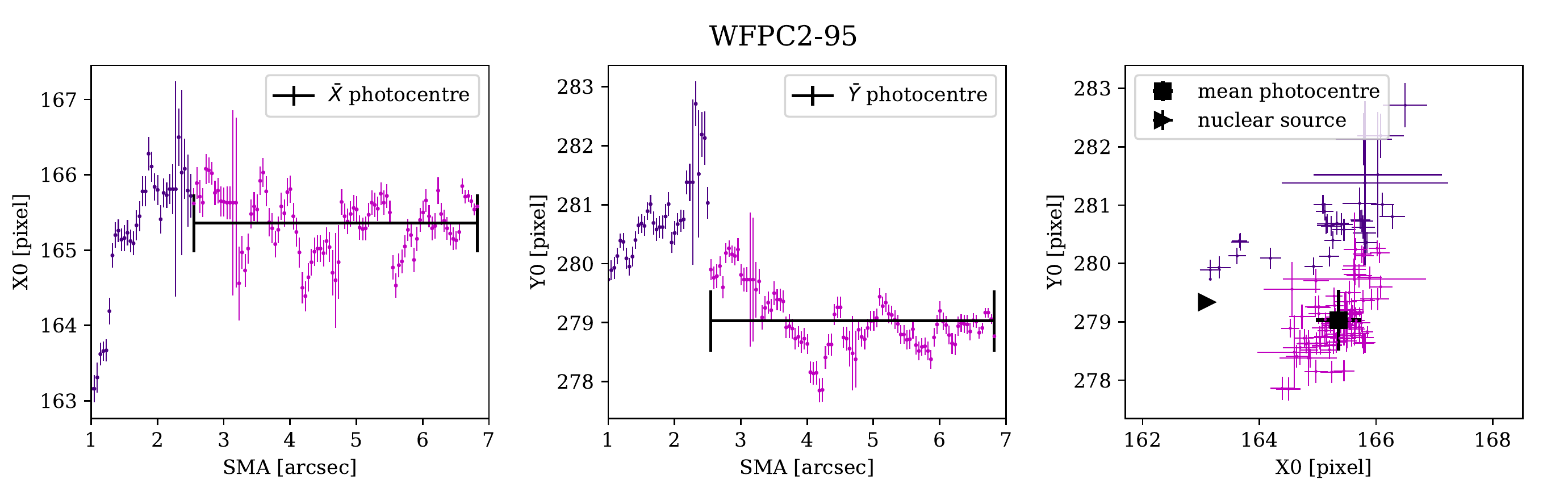}}\vspace{0mm}
\subfigure{\includegraphics[trim = 25mm 84mm 25mm 10mm, clip,width=1\textwidth]{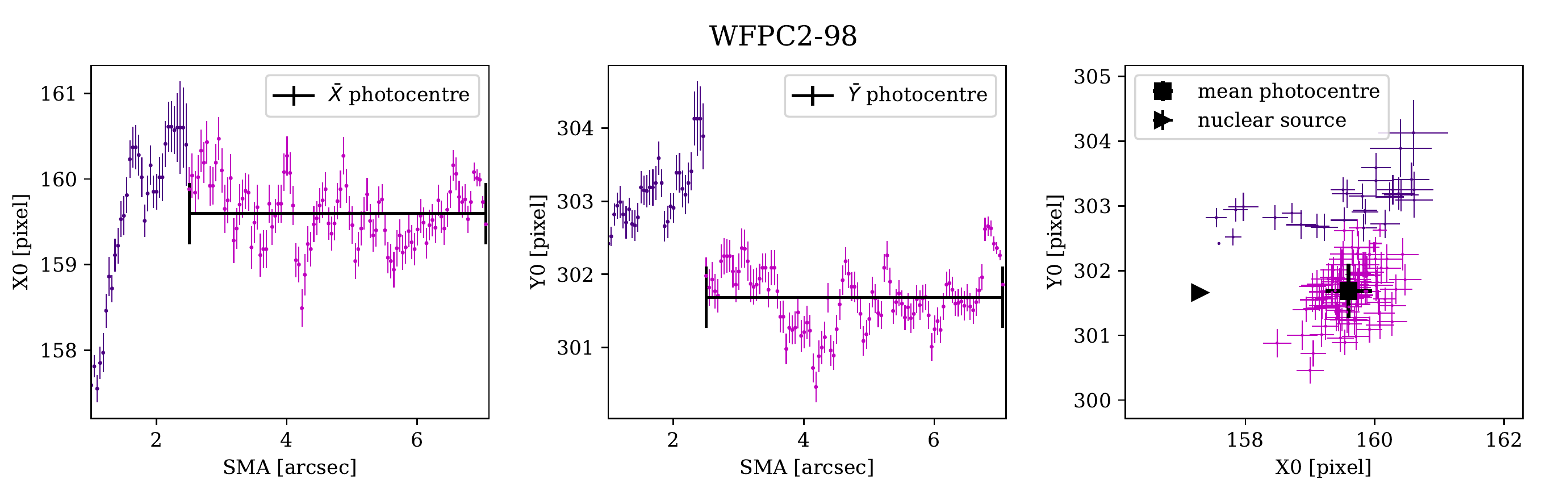}}\vspace{0mm}
\subfigure{\includegraphics[trim = 22mm 84mm 25mm 10mm, clip,width=1\textwidth]{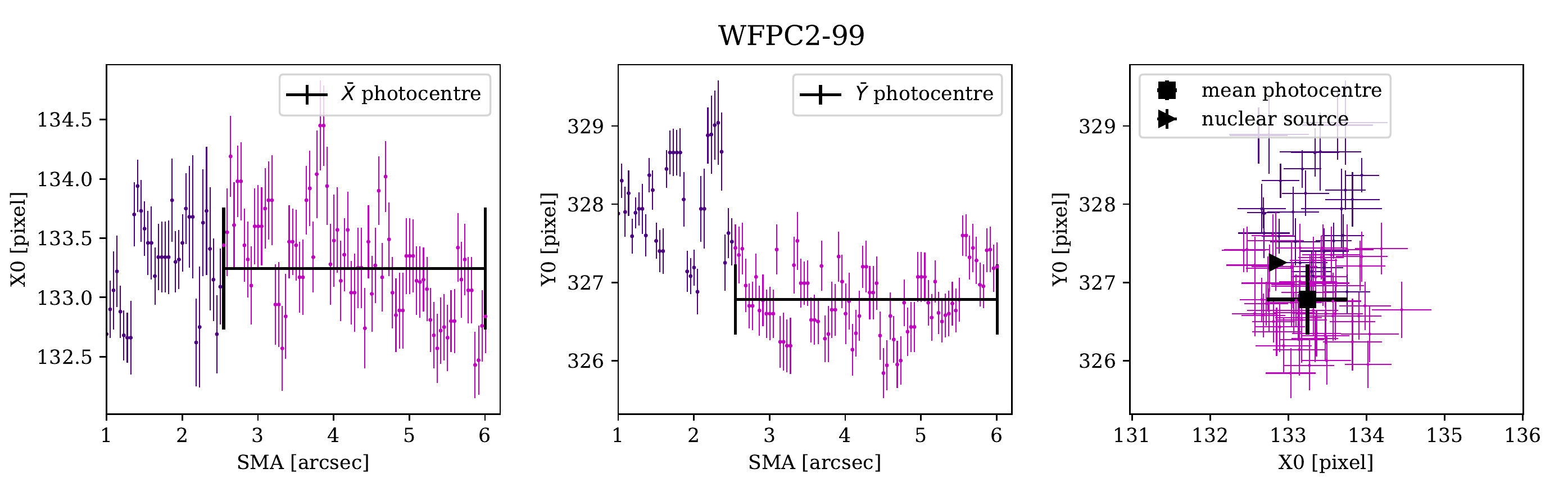}}\vspace{0mm}
\subfigure{\includegraphics[trim = 22mm 84mm 25mm 10mm, clip,width=1\textwidth]{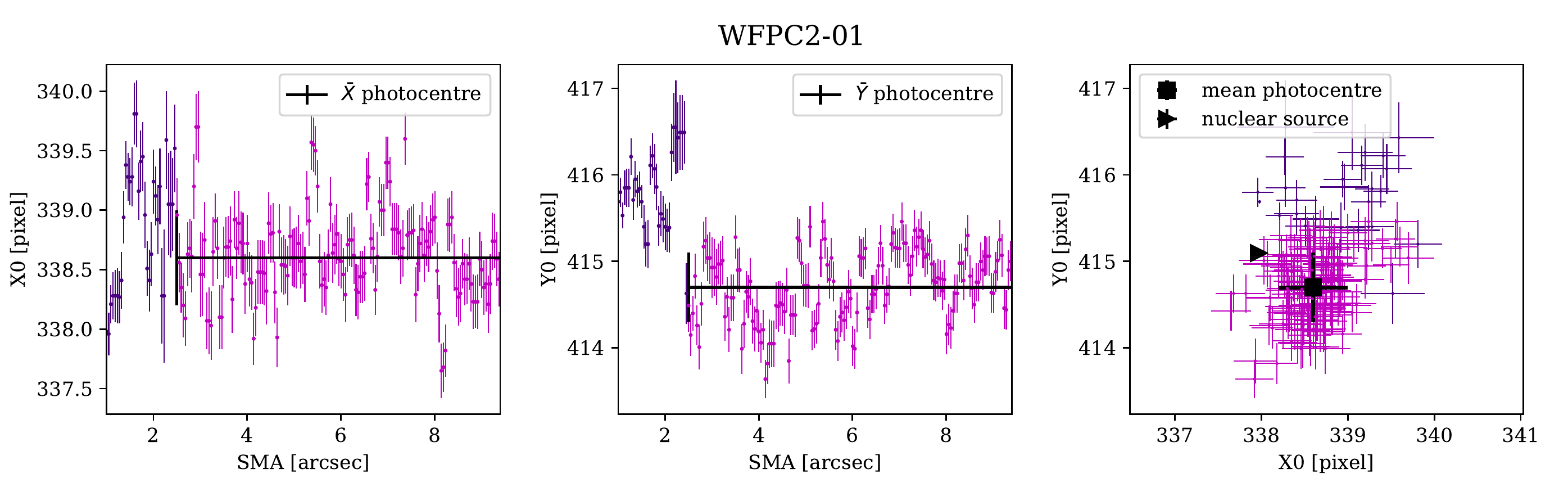}}\vspace{0mm}

\caption{First and second column: X and Y coordinates of the individual isophote centres as a function of ellipse semi-major axis (SMA). Third column: isophote centres and nuclear source positions. The mean isophotal centre positions derived from $\approx$2\arcsec are included. The maximum ellipse SMA is which fits in the available field of view but never larger than the break radius of M87 nuclear region, $r= 9.41$\arcsec.}
\label{phot1}
\end{figure*}
 
 \begin{figure*}
\centering
\subfigure{\includegraphics[trim = 25mm 84mm 25mm 10mm, clip,width=1\textwidth]{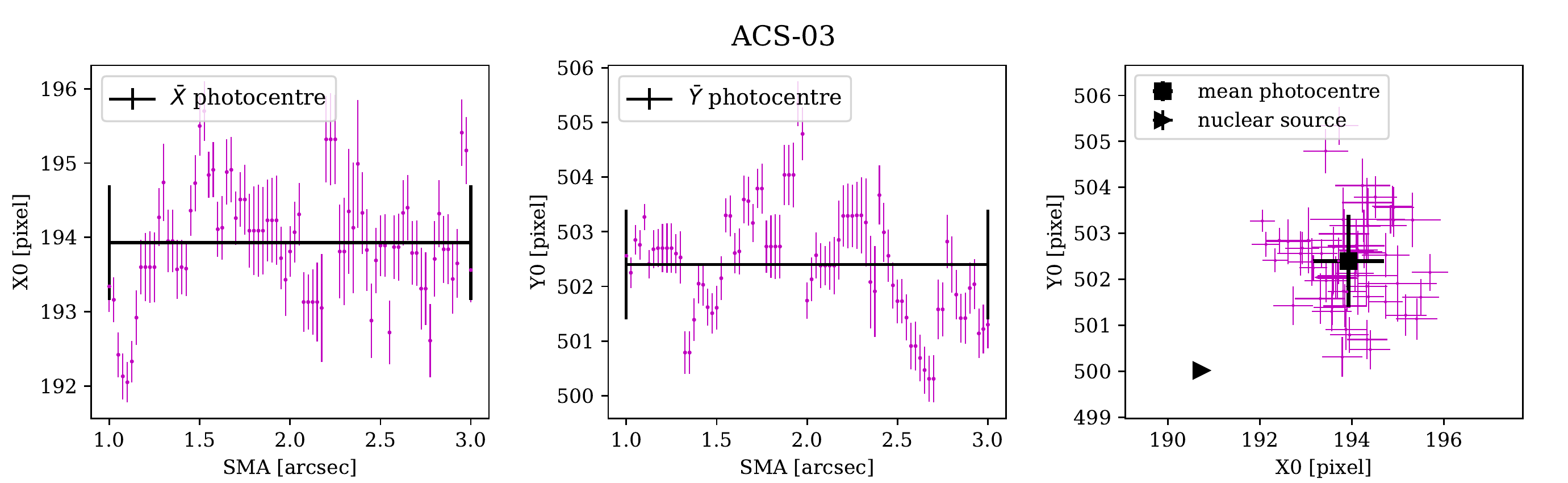}}\vspace{0mm}
\subfigure{\includegraphics[trim = 25mm 84mm 25mm 10mm, clip,width=1\textwidth]{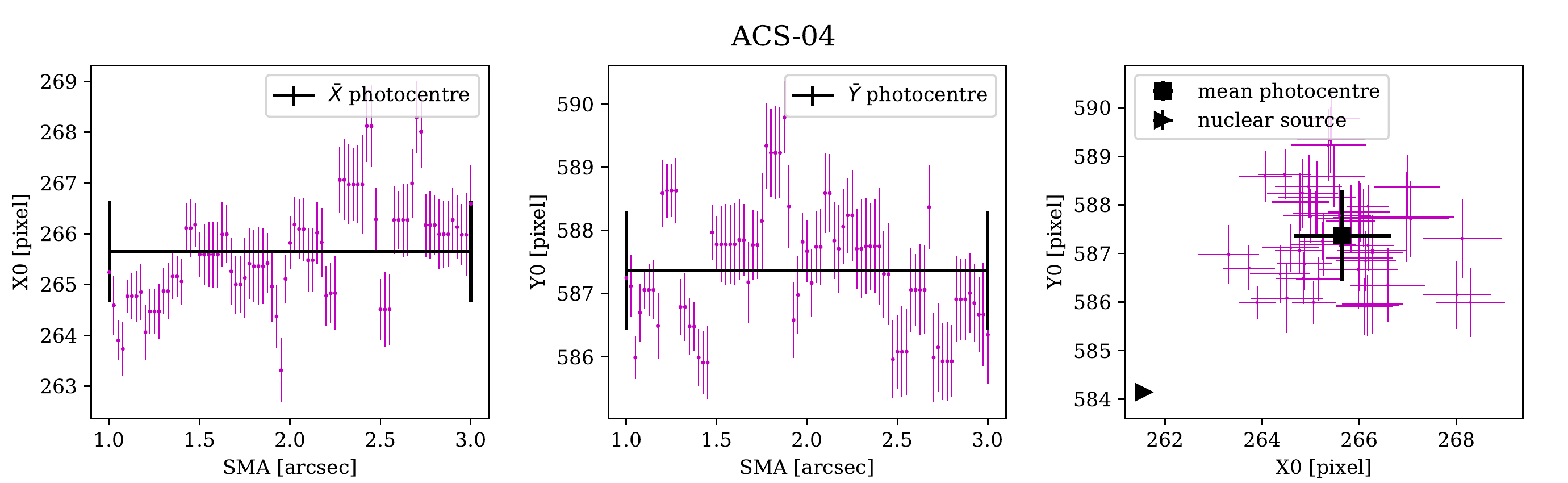}}\vspace{0mm}
\subfigure{\includegraphics[trim = 25mm 84mm 25mm 10mm, clip,width=1\textwidth]{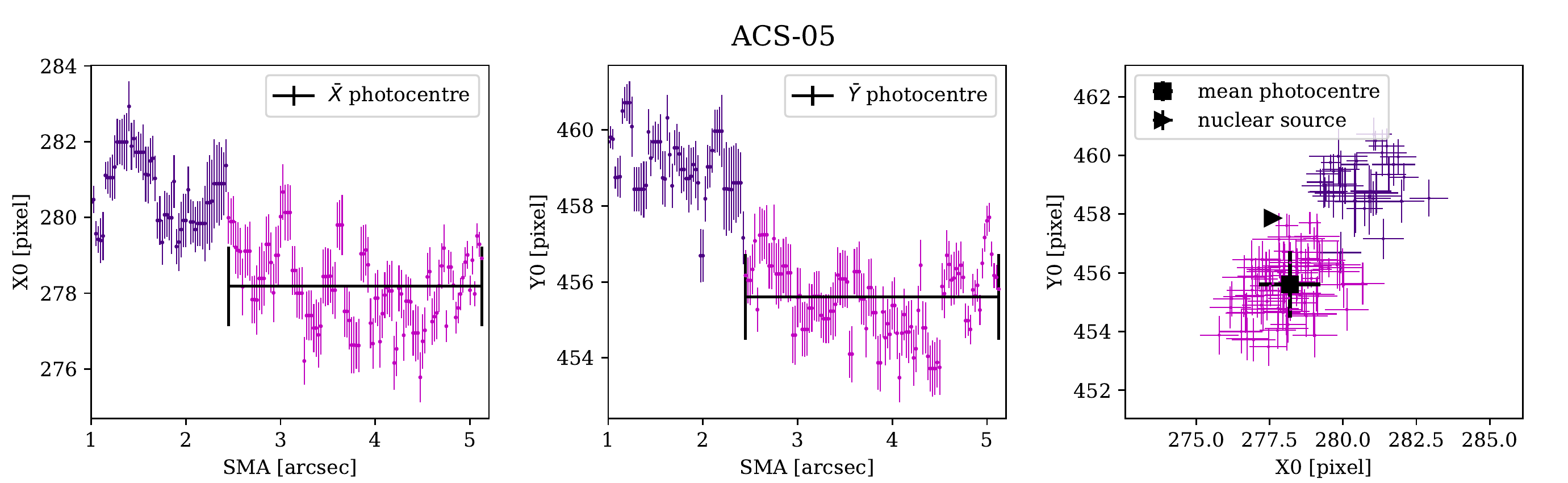}}\vspace{0mm}
\subfigure{\includegraphics[trim = 25mm 84mm 25mm 10mm, clip,width=1\textwidth]{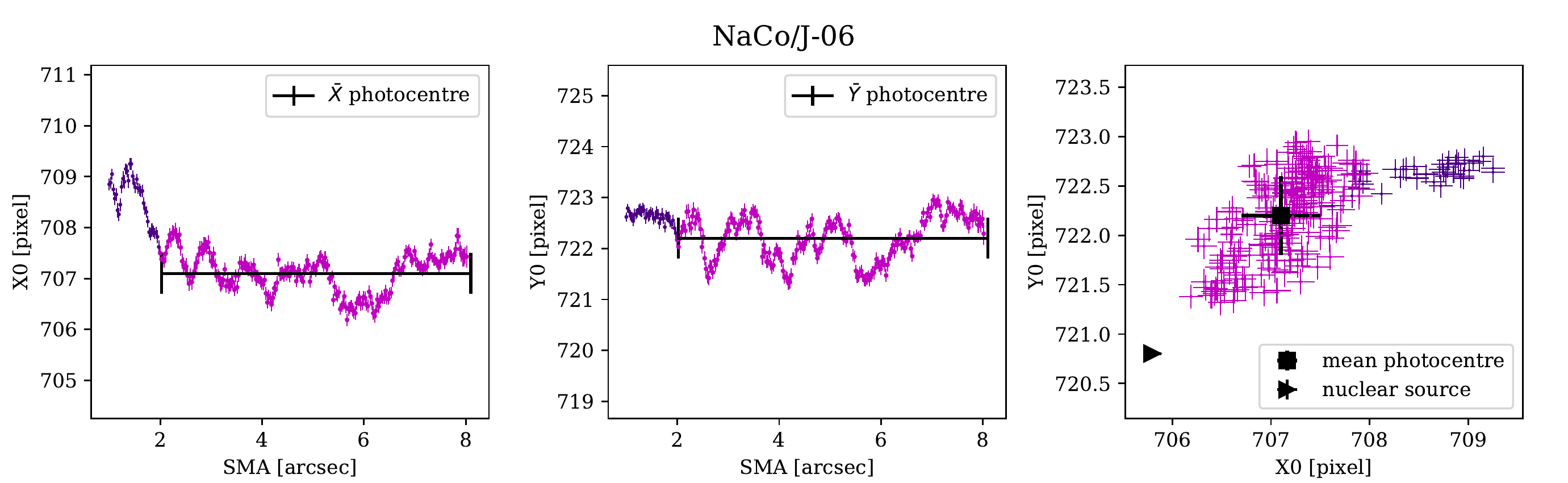}}\vspace{0mm}

\contcaption{First and second column: X and Y coordinates of the individual isophote centres as a function of ellipse SMA. Third column: isophote centres and nuclear source positions. The mean isophotal centre positions derived from $\approx$2\arcsec are included. The maximum ellipse SMA is which fits in the available field of view but never larger than the break radius of M87 nuclear region, $r= 9.41$\arcsec.}
\label{phot2}
\end{figure*}

 \begin{figure*}
\centering
\subfigure{\includegraphics[trim = 25mm 84mm 25mm 10mm, clip,width=1\textwidth]{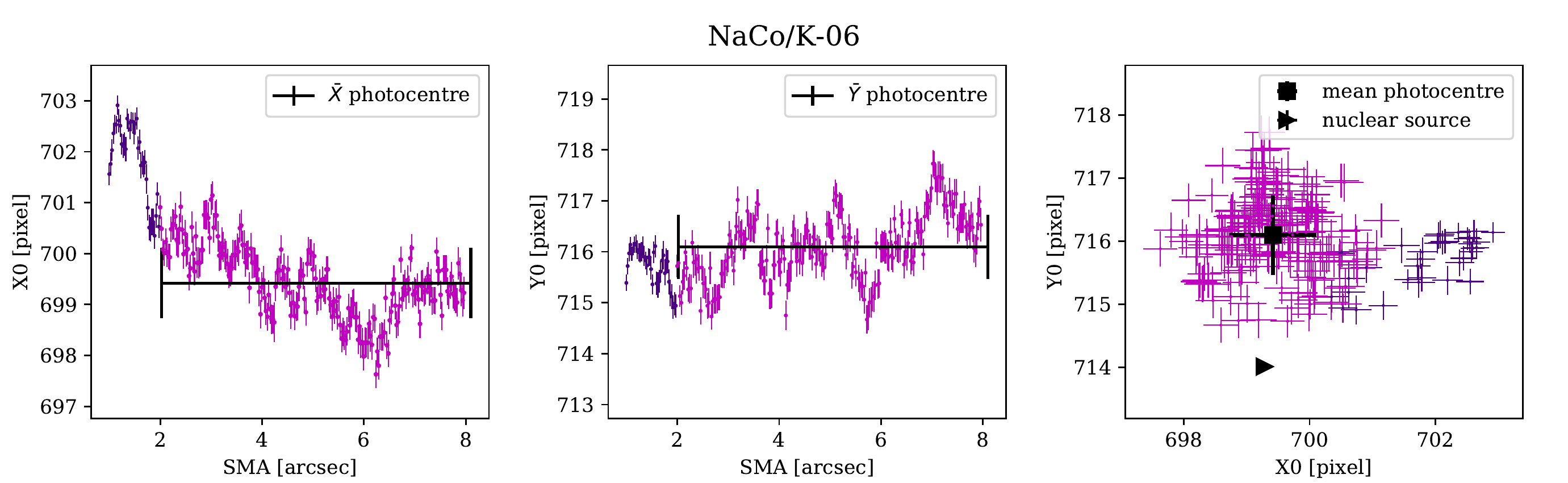}}\vspace{0mm}
\subfigure{\includegraphics[trim = 25mm 84mm 25mm 10mm, clip,width=1\textwidth]{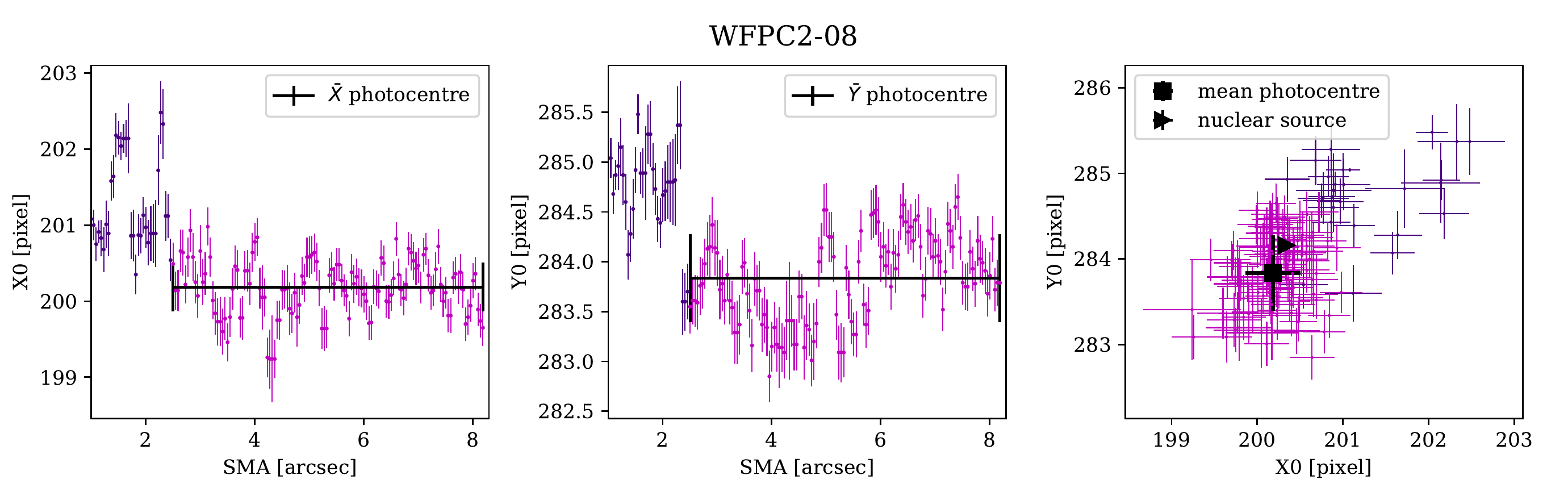}}\vspace{0mm}
\subfigure{\includegraphics[trim = 25mm 84mm 25mm 10mm, clip,width=1\textwidth]{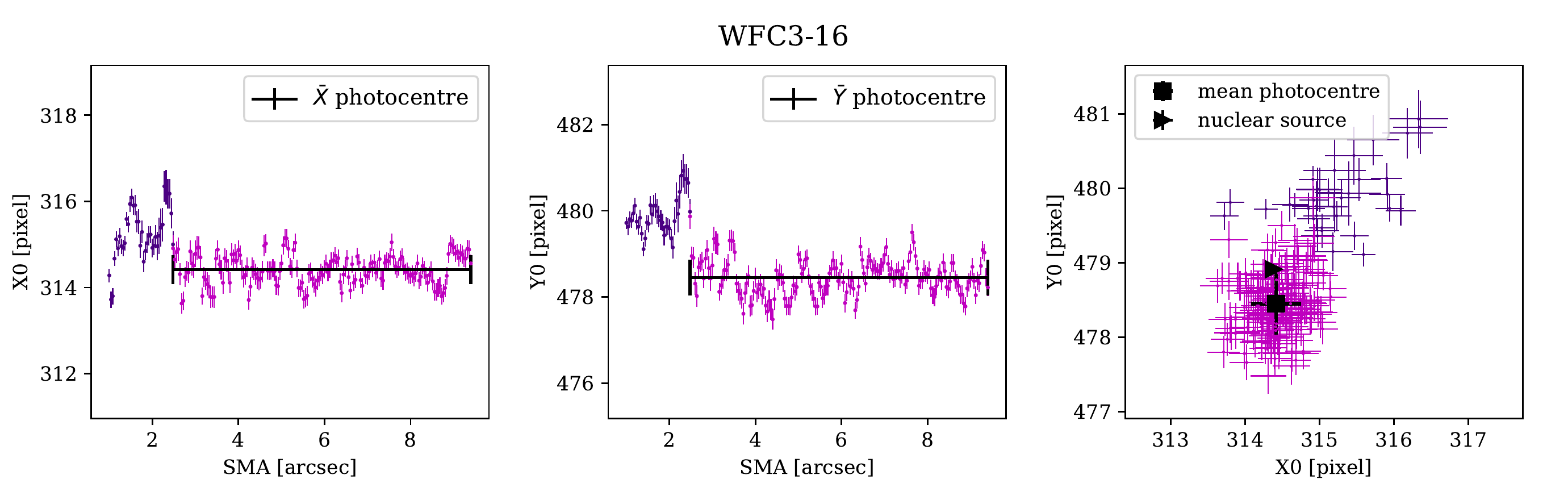}}\vspace{0mm}
\subfigure{\includegraphics[trim = 20mm 84mm 25mm 10mm, clip,width=1\textwidth]{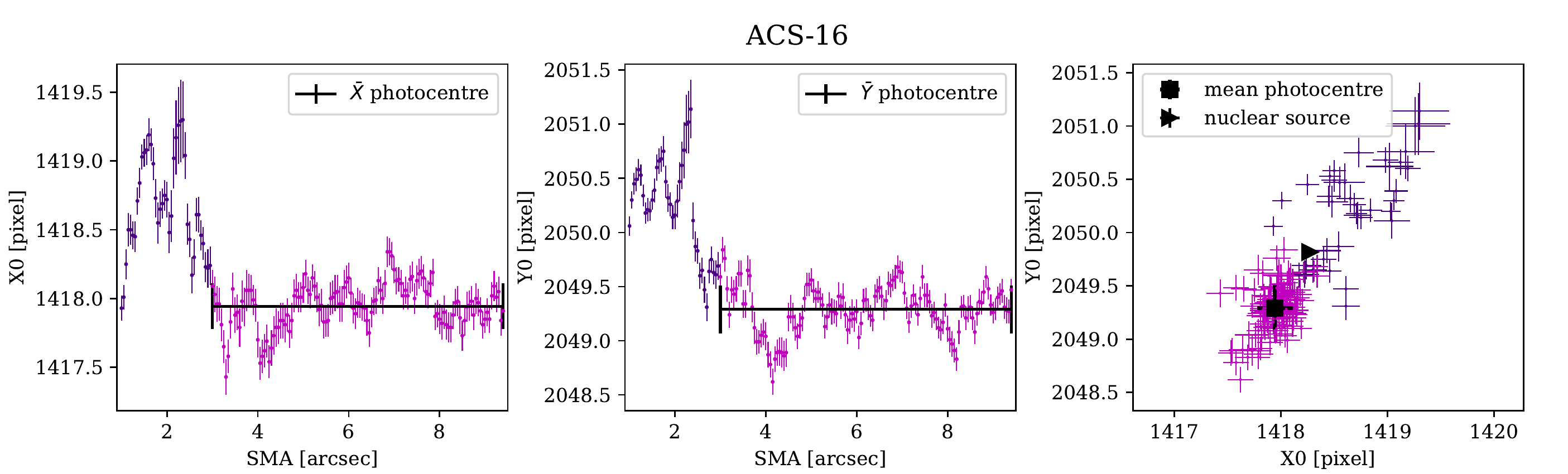}}\vspace{0mm}
\contcaption{First and second column: X and Y coordinates of the individual isophote centres as a function of ellipse semi-major axis (SMA). Third column: isophote centres and nuclear source positions. The mean isophotal centre positions derived from $\approx$2\arcsec are included. The maximum ellipse SMA is which fits in the available field of view but never larger than the break radius of M87 nuclear region, $r= 9.41$\arcsec.}
\label{phot3}
\end{figure*}
\end{document}